\documentstyle[epsf,psfig,natbib_mod]{mn}
\begin{document}

\title[Non-linear X-ray variability in XRBs and AGN]
{Non-linear X-ray variability in X-ray binaries and active galaxies}
\author[P. Uttley, I. M. M$^{\rm c}$Hardy and S. Vaughan]
{P. Uttley$^{1}$\thanks{e-mail: pu@milkyway.gsfc.nasa.gov},
I. M. M$^{\rm c}$Hardy$^{2}$ and S. Vaughan$^{3}$ \\
$^{1}$LHEA, Code 662, NASA Goddard Space 
Flight Center, Greenbelt, MD 20771, USA \\
$^{2}$School of Physics and Astronomy, University of Southampton,
Southampton SO17 1BJ \\
$^{3}$X-ray and Observational Astronomy Group, Department of Physics and Astronomy, University of
Leicester, Leicester LE1 7RH \\
}

\date{}

\maketitle
\parindent 18pt
\begin{abstract}
We show that the rms-flux relation recently discovered in the X-ray
light curves of Active Galactic Nuclei (AGN) and X-ray binaries (XRBs) implies that the
light curves have a formally non-linear, exponential form, provided the
rms-flux relation applies to variations on all time-scales (as it appears to).  This phenomenological
model implies that stationary data will have a lognormal flux distribution.  We confirm this result
using an observation of Cyg~X-1, and further demonstrate that our model predicts the existence 
of the powerful
millisecond flares observed in Cyg~X-1 in the low/hard state, and explains the general shape and 
amplitude of the bicoherence spectrum in that source.  
Our model predicts that the most variable light curves will show the most
extreme non-linearity.  This result can naturally explain the apparent non-linear variability
observed in some highly variable  Narrow Line Seyfert~1 (NLS1)
galaxies, as well as the low states observed on long time-scales in
the NLS1 NGC~4051, as being nothing more than extreme
manifestations of the same variability process that is observed in XRBs and
less variable AGN.  That variability process must be multiplicative (with variations coupled 
together on all time-scales) and cannot be additive (such as
shot-noise), or related to self-organised criticality, or result from completely
independent variations in many separate
emitting regions.  Successful models for variability must reproduce the observed
rms-flux relation and non-linear behaviour, which are more 
fundamental characteristics of the variability process than the power spectrum or 
spectral-timing properties.  Models where X-ray variability is driven by
accretion rate variations produced at different radii remain the most promising.
\end{abstract}


\section{Introduction}
The X-ray light curves of active galactic nuclei (AGN) and X-ray binary
systems (XRBs) are often dominated by strong flickering type
variability, which is aperiodic (noise-like),
e.g. \citet{mch88,vau03b,vdk95}.  The remarkable similarities
between various aspects of 
AGN and black hole XRB (BHXRB) variability suggest that
the same physical mechanism underlies the variability in both cases 
(\citealt{utt02,mar03,vau03a,mch04}).  Furthermore, similarities between
variability properties in neutron star and black hole XRBs \citep{wij99,bel02,utt01}, 
which show different X-ray spectra (e.g. \citealt{don03}) and presumably different X-ray
emission mechanisms, suggest that the same underlying variability process
is at work, regardless of the nature of the X-ray emission process. 

The flickering nature of AGN and XRB light curves makes
any physical interpretation of the variability difficult, but despite this
problem a variety of
models have been suggested to explain the variability
(e.g. \citealt{ter72,min94,pou99}).  Principally, such
models try to explain the shape of the power-spectral density function (PSD),
which as a first approximation can be treated as a broken or more gently
bending power-law (\citealt{bel90,now99,mch04})\footnote{Although recently it has been shown that
the low/hard state PSDs of XRBs are better represented as a sum of broad Lorentzians
\citep{now00,bel02,pot03}.}.  For example, additive
shot-noise models, where the light curve is produced by a sum of flares 
or `shots', seek to explain
the breaks in the PSD in terms of the maximum and minimum decay
time-scales in a distribution of shot-widths (\citealt{leh89,loc91}).  In
these models, the observed overall PSD shapes
are understood in terms of an assumed distribution of the shot widths,
however, in principle a shot-width distribution exists to fit any
noise process PSD, regardless of whether the shot interpretation
is physically meaningful,
so the models do not have much predictive or explanatory power \citep{doi78}. 
Models where
variability is associated with self-organised criticality (SOC) in the 
accretion flow \citep{min94} solve this problem by predicting a 
specific power-law PSD as a
natural outcome of the SOC process (\citealt{bak88,chr91}), 
but these models also require some 
modification in order to produce the observed PSD slopes \citep{tak95}. 

An important lesson of the work on shot-noise and SOC models is that
although the PSD is a very useful tool for quantifying variability (e.g. in
measuring characteristic time-scales), it has some limitations in 
distinguishing between models for
aperiodic variability.  This point is reinforced by the
fact that a common PSD shape is in fact {\it not} a defining
characteristic of the X-ray variability, because the PSD shape evolves 
over time. In particular, the PSD shape changes
dramatically between different spectral states in BHXRBs (e.g. \citealt{mcc03,pot03}).
 Other characteristics of the variability may provide stronger
model constraints. One recent observation which
strongly constrains models of variability is
the discovery of a correlation between the X-ray variability amplitude and the
flux in AGN and XRBs \citep{utt01}.  Specifically, the absolute (not fractional)
amplitude
of root-mean-squared variability increases linearly with the mean flux level, 
implying that a given source is in some sense more variable when it is brighter. 
This linear rms-flux relation is observed in all known
spectral states of the BHXRB Cyg~X-1, independent of PSD shape,
suggesting that the rms-flux
relation is a more fundamental characteristic of the variability than the PSD
shape \citep{gle04}.  Apparently linear
rms-flux relations are also observed in a variety
of AGN (\citealt{ede02,vau03a,vau03b,mch04}),
albeit at a lower signal-to-noise than in the excellent XRB data where
highly linear relations are observed in both BHXRBs and neutron star 
XRBs \citep{utt01,utt04}.

In \citet{utt01}, we argued that the observation of a
linear rms-flux relation in XRB and AGN X-ray light curves rules out
additive shot-noise models for the variability, because it implies that shorter
time-scale variations must somehow `know about' the behaviour of the source on
longer time-scales (or equivalently, the shorter and longer time-scale
variations are coupled together).  Additive shot-noise models treat the 
shots on all time-scales to be independent of one another and so cannot
produce this effect.  We originally
suggested that models should be considered where
the longer-term variations precede the short-time-scale variations,
e.g. as in the perturbed accretion-flow
model of \citet{lyu97} where accretion rate variations on different 
time-scales propagate inwards through the accretion flow to modulate 
the X-ray emission (see also \citealt{kin04}).  This model
is also strongly supported by the recent discovery that the aperiodic
variability carrying the rms-flux relation in the accreting millisecond
pulsar SAX~J1808.4-3658, is modulated by the 401~Hz pulsations, implying an 
origin at the neutron star surface so that the variability cannot be
produced in the corona, e.g. by magnetic flares, and is most likely 
associated with the accretion flow \citep{utt04}.

In this paper, we expand
on our original work to consider the
phenomenological implications of the rms-flux relation.  In particular, 
after discussing some important time-series definitions in Section~\ref{def},
we demonstrate in Section~\ref{rmsflux}
that the observed rms-flux relations imply that the
light curves $x(t)$ are formally {\it non-linear}, and can be generated
by a simple transformation from linear data: $x(t)=\exp[l(t)]$.  This latter 
relation also suggests that the fluxes follow a {\it lognormal} distribution.  
Thus, the rms-flux relation, non-linear behaviour and a lognormal
flux distribution represent three different aspects of the same
underlying process.
We compare the predictions of our 
phenomenological `exponential model' for variability with real data for 
Cyg~X-1 in Section~\ref{modcomp}, confirming that the fluxes do indeed 
follow a lognormal distribution, and showing that our model can explain
some of the observed consequences of non-linearity, such as the existence
of occasional powerful flares \citep{gie03} or the
amplitude and shape of the bicoherence function \citep{mac02}. In 
Section~\ref{discuss}, we discuss the implications of our results, which
impose strong constraints on any models for variability which seek to reproduce
the phenomenological behaviour discussed here. The trio of 
effects resulting from our phenomenological model
can explain a variety of observed AGN behaviour as
being part of the same variability process, and strongly
imply that successful models
for AGN and XRB variability must be `multiplicative'
and not additive (like shot noise or SOC models), or 
deterministic (like dynamical chaos).  We conclude with some advice for testing models
of AGN and XRB variability.

\section{Some definitions}
\label{def}
Before we examine the implications of the rms-flux relation for the nature of
the variability, we first consider some important time-series issues, and definitions
which will be used in the remainder of the paper.
We only cover these topics in a
cursory fashion here, but a much deeper discussion of the subject can
be found in a number of standard texts (e.g. \citealt{pri82,kan97}).

\subsection{Processes, systems and models}
Following standard definitions in time-series analysis
(e.g.\citealt{the94}) we note that an observed time-series (i.e. a
light curve) is a realisation of the underlying stochastic {\it
process} which is sampled by the observation.  The process is
generated by the physical {\it system} which produces the variability,
and a major goal of any time-series analysis is to determine the
nature of that system.  However, in practice it may only be possible
to determine a mathematical {\it model} which can reproduce the
observable properties of the process, and only relate that model to
the physical system using knowledge of the appropriate physics.  It is
important to remember that since the observed light curve is only a
realisation of the underlying process, a complete and/or accurate
statistical description of that process and the corresponding model
can be difficult to obtain from real data.  To demonstrate these
distinctions, consider an observed light curve with a doubly broken
power-law PSD shape.  The observer must first assume that the observed
PSD is a good representation of the PSD of the underlying process.
Next the observer may assume that a shot-noise model is suitable to
reproduce the process (based on the observed PSD, e.g. \citealt{leh89}).
Finally, the observer may interpret that shot-noise model in terms of
a physical system consisting of 
independent X-ray flares due to magnetic reconnection in a corona.

Note that, in the hypothetical example given above, steps of inference
are made at each stage in interpreting the data, which may not be
warranted given better data or a more complete statistical description
of the data.  For example, a fundamental but often unstated assumption
is that the underlying process is {\it stationary}, so that the statistical
properties of the process (i.e. its `moments') remain constant with time. This strict-sense 
definition of time-series stationarity is often referred to as {\it strong 
stationarity}.  In practice, a less restricted definition of `weak stationarity' is used,
often referred to simply as stationarity, since this is the form of stationarity most 
commonly assumed and we use this terminology here also.
For a weakly stationary process, only the first two moments are constant (mean, and autocovariance, 
i.e. variance and autocorrelation function, or equivalently the PSD).  
Red-noise light curves, which are
realisations of the underlying stochastic process, can strictly only be
considered as {\it weakly non-stationary} in the sense that they 
have a mean and variance which
change with time (due to the statistical fluctuations inherent in the noise 
process).  However, it is often assumed that the underlying process is stationary,
since on long time-scales the red-noise PSD should flatten (to power-law indices $>-1$)
in order to preserve a
finite total variance, and the light curve mean and variance will asymptotically
converge on the true mean and variance of the process (the process is said to be
asymptotically stationary).
Similarly, if light curves are much longer than the
longest variability time-scales produced by the underlying stationary
process, those light
curves can themselves be considered as stationary, since their
statistical properties will approximate closely those of the underlying 
process.

\subsection{Linearity and non-linearity}
We are now in a position to formally define linearity and
non-linearity in the context of time-series.  A linear process is one
that can be described by a model whose output (i.e. the process) is linear with respect to
the inputs to the model, e.g. so that multiplying the inputs by a
constant multiplies the output by the same constant.  For example,
following the definitions of \citet{pri82} a
general linear model to produce the linear process $L_{i}=L(t_{i})$, consisting of
discrete time steps $t_{i}$, $i=1,2,3 \ldots$ is:
\begin{equation}
L_{i}=\sum^{\infty}_{j=0} g_{j}u_{i-j} \label{lindef}
\end{equation}
where $u_{i}$\footnote{Following standard mathematical notation,
$i$ here and elsewhere in the paper
is a dummy index and so is interchangable with $j$ or $i-j$},
is a sequence of independent random variables,
so that at each time step
the value of the process $L_{i}$
(the `flux') is given by the sum of random variables from step $i-\infty$
to step $i$, each multiplied by the corresponding element in 
the sequence $g_{i}$, which essentially denotes the `memory' in the
time-series, i.e. how correlated the data point $L_{i}$ is with
the data at previous times, $L_{i-j}$.  To give three simple examples, if $g_{i}=0$ for
$i\neq0$, the time-series would be completely uncorrelated, white-noise
data; if $g_{i}$ is a constant $>0$ for all $i$, the data would be correlated on
all time-scales, representing a random walk form of red-noise data; if
$g_{i}>0$ for small $i$, becoming 0 at larger values, the
data would be correlated on short time-scales only, i.e. its PSD would flatten to
zero-slope at low frequencies.
Note that the sum of the squares of the $g_{i}$ co-efficients is proportional
to the total variance of the light curve (see \citealt{pri82}, Chapter~10.1.1).

By contrast, a non-linear process is one that does not conform to a 
linear model.  For example, the process $X_{i}$ generated by a model called
the `Volterra expansion' \citep{pri82} is
non-linear in the inputs because of additional higher order multiplicative
terms in the model:  
\begin{eqnarray}
X_{i} & = & \sum^{\infty}_{j=0} G_{j}u_{i-j}
+\sum^{\infty}_{j=0}\sum^{\infty}_{k=0}G_{jk}u_{i-j}u_{i-k} +
\nonumber \\
& & \sum^{\infty}_{j=0}\sum^{\infty}_{k=0}\sum^{\infty}_{l=0}
G_{jkl}u_{i-j}u_{i-k}u_{i-l}+ ...
\label{volterra}
\end{eqnarray}
where the $u_{i}$ are strictly independent random variables and the
$G$ coefficients of the expansion carry out a similar role to the $g$
coefficients in Equation~\ref{lindef}.  If the higher order terms
$G_{jk}, G_{jkl}, \ldots$ are all zero then the equation reduces to 
Equation~\ref{lindef} and the process is linear.
Note that, as pointed out by
\citet{sca97}, it is not strictly the time series or process which is
non-linear, but rather the model which describes it.  Therefore it
is possible to observe light curves which may have the appearance of
non-linearity but can be described by a linear model and hence (for
the purposes of definition) can be considered linear. 
For example, in recent years, evidence has been claimed for
non-linearity in the large amplitude X-ray variability of a number of
Seyfert galaxies (e.g.
\citealt{lei97,gre99,gli02}), but due to the limited data it is
difficult to reject the hypothesis that the data are linear
but non-Gaussian (see discussion in \citealt{lei99}).
\footnote{The much more
rapid variability observed in XRBs provides much better statistics however, 
and measurements of the `bicoherence' 
have conclusively detected non-linearity in light curves of the 
black hole candidates Cyg~X-1 and GX~339-4\citep{mac02}}.

Typically, a stationary {\it Gaussian} process, that is, a process 
with a Gaussian distribution of
$L_{i}$, can be produced by linear models which involve the addition of very many
small elements, e.g. certain shot-noise models.  Gaussianity then
follows from the central limit theorem.  Processes with non-Gaussian
distributions of $L_{i}$ can be produced by linear models
where only a small number of additive elements are involved (provided
their flux distributions are non-Gaussian), or from, e.g. a Poisson
process (see \citealt{lei99} for an example).  But non-Gaussian processes
can also arise when the model is non-linear, e.g. multiplying elements
together, rather than adding them, as we shall see below.
For describing observed light curves,
it is important to be careful to distinguish models which are non-Gaussian and linear
from models which are non-Gaussian and non-linear (e.g. see \citealt{the92}).

\subsection{Lognormal processes}
One distribution of time-series data which is commonly found in
Nature is the {\it lognormal} distribution \citep{ait57,cro88}.  The lognormal
distribution can be thought of as the analogue of the normal/Gaussian
distribution for multiplicative rather than additive processes.  For
example, consider a stationary process $X$ which is the result of
$N$ random subprocesses $x_{i}$ which multiply together, so that
$X=\prod_{i=1}^{N} x_{i}$.
Therefore the logarithm of $X$ is the sum of the logarithms of the
individual $x_{i}$.  As $N\rightarrow\infty$ then (provided
the $x_{i}$ are independent and identically distributed, but
regardless of the shape of that distribution) the distribution of the sum of the
logs of $x_{i}$, $\log[X]$ must
approach a Gaussian distribution (by the central limit theorem).
Therefore a process produced by multiplication of many independent
processes will have a lognormal distribution, where the distribution
of $\log[X]$ is Gaussian.  A general univariate form of the lognormal
distribution is the three-parameter lognormal distribution:
\begin{equation}
f(x;\tau,\mu,\sigma)=\frac{1}{\sigma \sqrt{2 \pi} (x-\tau)} \exp \frac{-[\log(x-\tau)-\mu]^{2}}{2\sigma^{2}}
\label{lognormeqn}
\end{equation}
where $\tau$ is a `threshold parameter' representing a
lower limit on $x$ (e.g. caused by some constant offset which is
additive to $x$) and $\mu$ and $\sigma^{2}$ in this case represent the
mean and variance of the distribution of $\log[x-\tau]$.  The lognormal
distribution has a long history in describing a wide range of
phenomena such as economic data, population statistics, or size distributions such
as cloud sizes and grain sizes in sand (e.g. see
\citealt{cro88} for an overview).  The ubiquity of lognormal
distributions in Nature results from the fact that many
natural processes are multiplicative (e.g. increases in populations,
the random splitting of clouds or grains of sand).  
In the context of this work, lognormal statistics have been found to
apply to the fluences of events in shot-fitting models of GRB and X-ray
binary variability data \citep{qui02,neg02}, as well as the flux 
distribution of the extremely variable NLS1 IRAS~13224-3809 \citep{gas03}.

\section{The rms-flux relation and non-linearity}
\label{rmsflux}
\subsection{The nature of the rms-flux relation}
The absolute rms amplitude of variability $\sigma_{\rm rms}$ of a time series of
$N$ data points, $X_{i}$, is defined as the square-root of the light curve
variance, i.e.:
\begin{equation}
\sigma_{\rm rms} = \sqrt{\frac{1}{N-1}\sum_{i=1}^{N} (X_{i}-\overline{X})^{2}}
\end{equation}
where $\overline{X}$ is the mean of the series.
For weakly non-stationary time series it is typically assumed (e.g.
implicitly by most
simulation methods; \citealt{leh89,tim95}) that
the $\sigma_{\rm rms}$ measured from individual
segments of the time series
is not constant but varies randomly about
some mean value (determined only by the underlying power spectrum).
However, the observed X-ray light curves of AGN and XRBs are not merely weakly non-stationary
in this
sense: the $\sigma_{\rm rms}$ of segments of the light curve varies randomly
about a mean value which scales linearly with the flux of the segment
\citep{utt01}.  We say that the light curves display a
{\it linear rms-flux relation}.  The rms-flux relation
is most convincingly demonstrated in XRBs, where it can be probed
on short time-scales with high significance (e.g. \citealt{gle04,utt04}).
Recent studies of X-ray light curves of the NLS1 Ark~564 \citep{ede02},
NGC~4051 \citep{mch04} and the Seyfert~1 AGNs MCG-6-30-15 and Mrk~766
\citep{vau03a,vau03b},
also show clear linear rms-flux relations in these
AGN\footnote{We note here that a number of other authors have investigated
the relationship of variability amplitude with flux in various AGN, with
varying results (\citealt{nan01,dew02}).  
However, as none of these authors took
account of the random variations in variability amplitude inherent in
weakly non-stationary noise processes
(e.g. see discussion in \citealt{vau03b}) it is difficult to assess
the significance of these results.}.
Therefore, the linear rms-flux relation may be fairly
ubiquitous in AGN and XRBs.

\subsection{Time-scale-dependence of the rms-flux relation}
\label{rmstimescale}
According to Fourier theory, any given time series can be decomposed into a set of sinusoidal signals
which represent the various time-scales of variability in the time series.  For an infinitely
long stochastic time series there are (in general) an infinite number of frequency components, although
for stationary or weakly non-stationary
time series, which have finite variance, the amplitudes of most of these components will be negligible
and the time-scales of significant variability will be concentrated into a certain range. 
Any {\it linear} stochastic time series (e.g. aperiodic variability)
can be synthesised by summing sine waves with 
random phases ($\phi$ uniformly distributed between 0 and $2 \pi$) and suitably
drawn amplitudes $A_i$. The amplitudes of the components $A_i$
can be determined from the power spectrum (e.g. see \citealt{tim95}).
Thus a linear time series (with zero mean) can be written:
\begin{equation}
\label{eqn:sum}
l(t) = \sum_{i=1}^{\infty} A_i \sin ( 2 \pi \nu_i t + \phi_i )
\end{equation}
This time series has no dependence between rms and flux.  Note that light curves 
generated by this method are Gaussian.  Non-linear light curves may be generated in a similar
way if the phases of the different components are not independent, but are correlated
with one another (e.g. see later discussion in Appendix~\ref{bicohapp}).

The aperiodic light curves of AGN and X-ray binaries show variations over a broad
range of time-scales, as is demonstrated by their broad, continuum-like PSDs, which contain
significant power over at least a decade range in frequency.  When measuring the rms-flux
relation, we are effectively considering three
ranges of time-scales (e.g. see \citealt{utt01}).  First, we can choose
the length, $T_{\rm seg}$
of the individual light curve segments which we use to measure the rms.  
Secondly, we can 
measure the rms over a specified range of time-scales (or equivalently,
frequencies) within
a light curve segment, by measuring the PSD of each segment and integrating the PSD 
over the
frequency range of interest to obtain the variance, taking the square root to obtain the
rms contributed by variations over that range of frequencies (time-scales).  Finally,
we plot the rms versus flux measured from each segment,
and hence examine the response of rms to flux variations on time-scales
$>T_{\rm seg}$.  In principle, we can isolate the response of rms on any given range
of time-scales to flux variations on any range of longer time-scales, by carefully
choosing the length of $T_{\rm seg}$ and the frequency range used to measure rms.
E.g. One can measure the response of 2-20~Hz rms to variations on time-scales $>100$~s
by choosing $T_{\rm seg}=100$~s and integrating over only 
the 2-20~Hz range of the resulting light curve segments.

Such an approach can be used to examine whether, e.g. 2-20~Hz rms responds only to
variations on time-scales $<10$~s, and in turn by changing the length of time segments,
and the frequency range integrated over, whether
flux variations on time-scales $>10$~s
themselves show a linear rms-flux relation.  For example, consider the case
where there are only two components
to the light curve, a fast variability component, and a slower component, which
modulates the amplitude of the fast variations but is not itself coupled
to variations on any longer time-scale (i.e. the slow component 
can vary in an arbitrary way and be treated as a simple `volume control'
for the amplitude of variations of the fast component, without any additional constraints
on its own amplitude of variability). Then
we will see a relation between the rms amplitude of the fast component and the
long time-scale flux variations due to the slow component.  However, if we
then measure the rms amplitude of variations of the slow component and correlate
them with the flux variations of the same component on
even longer time-scales we will not see a linear rms-flux relation.

\begin{figure}
\begin{center}
{\epsfxsize 0.9\hsize
 \leavevmode
 \epsffile{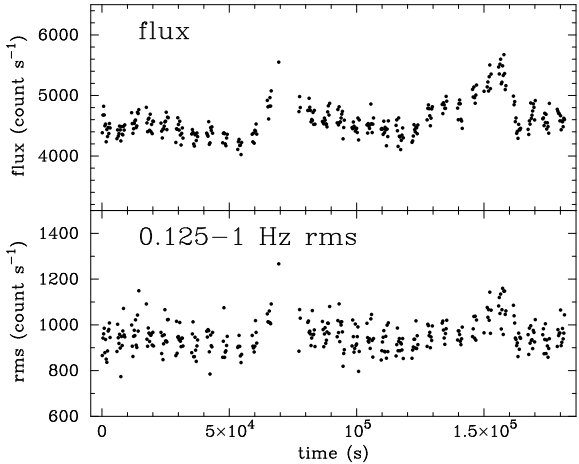}
}\caption{Flux and rms light curves of Cyg~X-1.  Measurements are made
from 256~s segments of the 2-13~keV light curve.  The rms is measured
over the 0.125-1~Hz band of the power spectrum obtained for each
segment.  Note there is some additional scatter in rms which is not in the flux
light curve, and is caused by random variations
due to the noise process.} \label{rmslc}
\end{center}
\end{figure}

In \citet{utt01}, we showed
that the rms-flux relation in Cyg~X-1 operates over at
least a decade range in time-scales
(from seconds, to tens of seconds).  We can test the time-scale dependence
of the rms-flux relation on longer time-scales by plotting the rms measured
in small segments as a function of time, i.e. we can make an
`rms light curve'.  Fig.~\ref{rmslc} shows the rms light curve of Cyg~X-1 measured
by the {\it Rossi X-ray Timing Explorer} ({\it RXTE}) in December 
1996\footnote{Proposal Number 10236}, compared
to the conventional flux light curve.  Clearly the variations
in rms are tracking the flux variations on time-scales of
hours.  Because the rms variations can be plotted as a time series in
their own right, it is possible to make a power spectrum of those
variations and so compare with the power spectrum of the flux light
curve.  \citet{gle04} have used this approach to show that
the PSD of short-time-scale (seconds) variations in rms is similar to
the PSD of the flux light curve, providing further evidence that the
rms variations follow the flux variations over a wide range of
flux variability time-scales.  \citet{gle04} have also
demonstrated that the high
frequency (1-32~Hz) rms in Cyg~X-1 responds linearly to flux changes
even on time-scales of months.  Therefore, it seems likely that the rms-flux
relation applies over a very broad range of time-scales, such that for any given
time segment duration we choose to measure the rms for, we will find a linear 
correlation between the rms amplitude and the flux of the segment.  

\begin{figure*}
\begin{center}
%\vspace{-0.5cm}
\hbox{
\psfig{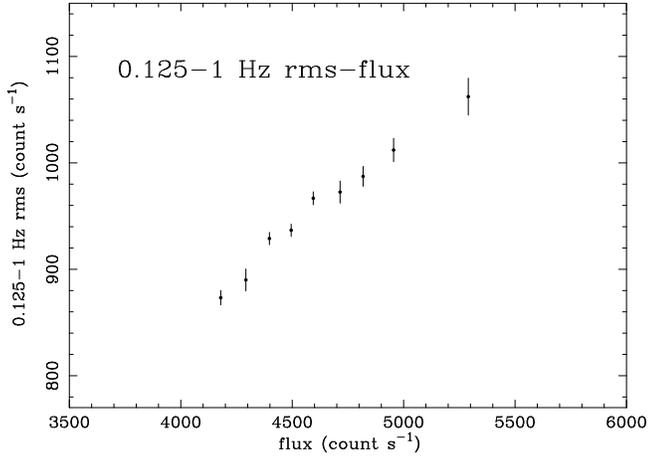}
\hspace{0.3cm}
\psfig{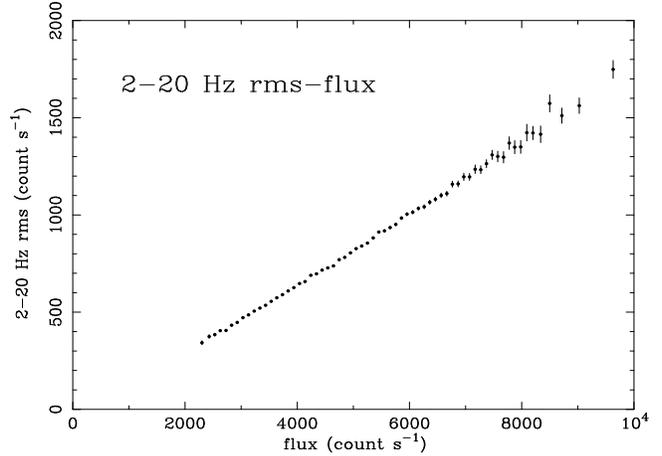}
}
\vspace{-0.5cm}
\end{center}
\caption{Cygnus~X-1 December 1996 2-13~keV rms-flux relations.  The left-hand panel shows the rms-flux
relation made by binning the 0.125-1~Hz rms measured in 256~s segments shown in 
Fig.~\ref{rmslc} according to the flux (note that due to the narrow range of fluxes,
the axes do not begin at zero).  The right-hand panel shows the rms-flux relation of the same
observation but with rms measured in the 2-20~Hz range for 1~s segments, prior to binning according
to flux.}
\label{dec1696rmsflux}
\vspace{-0.3cm}
\end{figure*}

\begin{figure}
\begin{center}
{\epsfxsize 0.9\hsize
 \leavevmode
 \epsffile{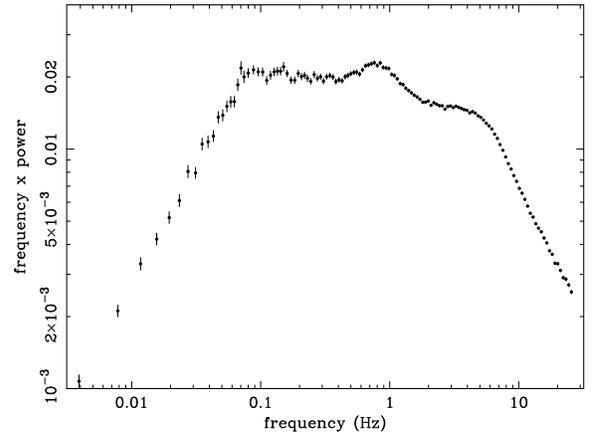}
}\caption{2-13~keV PSD of Cyg~X-1 in the low/hard state on 1996 Dec 16, plotted
in units of frequency$\times$power, so that a flat top corresponds to
a $1/f$ PSD shape.  Power units are $(rms/mean)^{2}$/Hz.} \label{dec1696psd}
\end{center}
\end{figure}

We illustrate
this point in Fig.~\ref{dec1696rmsflux}, which shows rms-flux relations
for the Cyg~X-1 data set shown in Fig.~\ref{rmslc}, measured for different segment
lengths, and hence dominated by variations on different time-scales.  Note that 
because the bulk of variability originates between $\sim0.1$~Hz and 1~Hz 
(as demonstrated in the PSD which we show in Fig.~\ref{dec1696psd}), the 
flux variations of the 1~s segments used to measure the 2-20~Hz rms-flux relation 
are dominated by variations in the 0.1-1~Hz range.  However, the 0.125-1~Hz
rms-flux relation clearly demonstrates that these variations themselves show a
linear rms-flux relation, i.e. the simple `volume control' model doesn't apply,
because the flux variations driving the 2-20~Hz rms variations themselves show an
rms-flux relation.

For the purposes of this paper, we will make the assumption, extrapolated
from the previous results described here, that Cyg~X-1 shows
a linear rms-flux relation on {\it all} time-scales.  By that, we mean that whatever
choice we make for the length of $T_{\rm seg}$ and the frequency range we use to 
determine rms, we will always see a linear rms-flux relation.  We will derive
in the next section a simple mathematical model for the variability which follows
from this assumption, and show in subsequent sections that this model
does indeed seem to explain many aspects of the data.

\subsection{The exponential form of AGN and XRB light curves}
\label{expderiv}
We now consider a simple mathematical model to approximate the rms-flux behaviour observed
in real light curves.  
In essence the linear rms-flux relation requires that the amplitude of
short time-scale variations is modulated by longer time-scale
trends in the data. The rms-flux correlation can thus be viewed
as an effect of amplitude modulation similar to that 
commonly encountered in e.g. radio communications.
To illustrate how amplitude modulation naturally leads to a linear
rms-flux relation, consider the modulation of a set of sine waves. 
Variations at a given frequency ($\nu_i$) are represented by a sine 
wave with unit mean and peak to trough amplitude less than twice the 
mean:
\begin{equation}
 f_i(t) = 1 + A_i \sin ( 2 \pi \nu_i t + \phi_i ) 
\end{equation}
where $0 < A_i < 1$ ensures that the wave is always positive,
and $\phi_i$ is the phase. It is simple to see that by multiplying
a high frequency sine wave (with frequency $\nu_1$) by another of
lower frequency ($\nu_2$), the amplitude of the high frequency
oscillations will be modulated by the low frequency oscillations.
The result will be a linear scaling between high frequency rms
amplitude measured in a time segment and the mean `flux' in that segment 
-- i.e. a linear rms-flux relation. 
This contrasts with the situation where the two sine waves
are added instead of multiplied together. In this case there
is no amplitude modulation and no dependence of rms upon flux. 

In the previous section, we noted that the rms-flux relation
is observed over a broad range of time-scales.  Of course, it is not possible to say for certain
that the rms-flux relation applies on {\it all} 
time-scales, e.g. that the rms responds to very long 
time-scale variations. 
However, since we are only considering the currently observable behaviour of XRBs and AGN,
we make the simplifying assumption that the linear rms-flux relation does apply to variations
on all time-scales.  We can represent this behaviour by making an
analogy with the sum-of-sines representation of a time series shown in Equation~\ref{eqn:sum},
to synthesise a general time series possessing
a linear rms-flux relation on all time-scales by extending the multiplicative sine
model:
\begin{equation}
\label{eqn:mult}
x(t) = \prod_{i=1}^{\infty} \{ 1 + A_i \sin ( 2 \pi \nu_i t + \phi_i ) \}
\end{equation}
To examine the properties of this model, we represent the sine components as 
$a_{i}(t)=A_i \sin ( 2 \pi \nu_i t + \phi_i )$, and
convert to a linear model by taking the logs, thus:
\begin{equation}
\label{eqn:logmult}
\log [x(t)]=\sum_{i=1}^{\infty} \log[1+a_{i}(t)]
\end{equation}
Since the phases of the individual sine wave components are uncorrelated, the $a_{i}$ are independent and
randomly distributed for a given time $t$.  If the underlying process is aperiodic and stationary, so
that the
power is spread out over many frequencies, then
no single frequency or small number of frequencies dominates the distribution of $\log[x(t)]$.  Under
these conditions we can apply the central limit theorem, finding that the distribution of $\log[x(t)]$ 
is Gaussian.  Therefore our model light curve $x(t)$ has a {\it lognormal} flux distribution.

Since $a_{i}(t)<1$, we can expand the logarithms on the right hand side and sum each 
term separately to give:
\begin{eqnarray}
\label{eqn:logmultexpand}
\log [x(t)] & = & \sum_{i=1}^{\infty} a_{i}(t) -\frac{1}{2}\sum_{i=1}^{\infty} [a_{i}(t)]^{2}
+\frac{1}{3}\sum_{i=1}^{\infty} [a_{i}(t)]^{3} \nonumber \\
& & - \sum_{n=4}^{\infty} - \frac{-1^{n}}{n}\left( \sum_{i=1}^{\infty} [a_{i}(t)]^{n}\right)
\end{eqnarray}
Since the even terms
are all postive definite they cannot be neglected
in comparison with the 1st order term, which we denote $l(t)=\sum_{i=1}^{\infty} [a_{i}(t)]$, since
it is a linear time series of the form given by Equation~\ref{eqn:sum}.  However, since the time-averaged
value of $l(t)$ is zero, we find that:
\begin{equation}
var[l(t)]=\langle \left(\sum_{i=1}^{\infty} a_{i}(t)\right)^{2} \rangle = 
\langle \sum_{i=1}^{\infty} [a_{i}(t)]^{2}\rangle
\end{equation}
where angle-brackets denote time-averages and the
latter equality holds because the time-averaged value of cross-terms $a_{i}(t)a_{j}(t)$
is also zero.  Hence the time-average of $\sum_{i=1}^{\infty} [a_{i}(t)]^{2}$
is equal to the variance of the linear time series. 
For a continuum process the number of sine components
in the signal approaches infinity, and so for the variance of the process to be finite, the amplitudes
$A_{i}\rightarrow0$, and hence $a_{i}\rightarrow0$.  The term
$\sum_{i=1}^{\infty} [a_{i}(t)]^{2}$ can be shown to be effectively
constant.  This is because, since the phases of the sine waves are independent of one another,
the variance of
this expression is equal to the sum of variances of the individual squared-sine terms (which can be 
simply evaluated).  Hence:
\begin{eqnarray}
\frac{var\left[ \sum_{i=1}^{\infty} [a_{i}(t)]^{2} \right]}
{\langle \sum_{i=1}^{\infty} [a_{i}(t)]^{2}\rangle^{2}} =
\frac{\sum_{i=1}^{\infty} [A_{i}^{4}]/8}{\left(\sum_{i=1}^{\infty} [A_{i}^{2}]/2\right)^{2}} \nonumber \\
= \frac{\sum_{i=1}^{\infty} [A_{i}^{4}]/8}{\left(\left[\sum_{i=1}^{\infty} \sum_{j=1}^{\infty} A_{i}^{2}
A_{j\neq i}^{2}\right] + \sum_{i=1}^{\infty} [A_{i}^{4}]\right)/4} .
\end{eqnarray}
The denominator is dominated by the sum of cross-terms (which are all positive): as the number of sine 
components tends towards infinity the ratio above tends to zero, hence the value of the 2nd order term
in Equation~\ref{eqn:logmultexpand} can be set to its time-averaged value, i.e. $var[l(t)]$, regardless of
$t$.
The
odd terms of 3rd order and higher have the same sign as the 1st order term and since $a_{i}\rightarrow0$,
they are always very small in 
comparison to $l(t)$.  Similarly, the even terms of 4th order or greater can be neglected in comparison
to $var[l(t)]$.  Therefore we can approximate $x(t)$ as:
\begin{equation}
x(t)\approx\exp\{l(t)-\frac{1}{2} var[l(t)]\}
\end{equation}
The variance term can be neglected, since it simply reflects a normalising constant.  Therefore
we can express the sine multiplication model for the rms-flux relation rather simply as
$x(t)\approx\exp[l(t)]$, i.e. aperiodic light curves with a linear rms-flux relation on all time-scales
can be produced by taking the exponential of a linear aperiodic light curve.
It is obvious that this model is non-linear with respect to the input linear `light curve' $l(t)$
(see also Appendix~\ref{nonlinproof} for a formal demonstration that the model is
equivalent to a Volterra expansion). 
Assuming that the input light curve is stationary,
it is important to note that the model itself generates a process which is stationary
in the sense described in
Section~\ref{def}, because it is a simple exponential transformation of a
linear process which is stationary.  Of course, if the assumptions underpinning the above derivation
break down, the exponential model will be an inappropriate representation
of the data.  For example, if the rms-flux
relation is only produced by a small finite number of multiplying components, the higher
order terms in Equation~\ref{eqn:logmultexpand} will become important, leading to deviations from
the model (and also deviations from the lognormal flux distribution, since if the number of contributing
components is small the distribution of $\log[x(t)]$ will no longer be Gaussian).

For completeness, we note here that the conclusion that the linear rms-flux relation implies that 
the logarithmically transformed data is Gaussian can be independently reached using
established methods for the transformation of uncorrelated non-Gaussian sample data (i.e. not
the time-series we measure here) into data with a Gaussian
distribution \citep{bar47,box64}.  Skewed, non-Gaussian data is {\it heteroskedastic}, that is 
the expectation value of its sample variance is not constant
and specifically, it may be a function of the mean of the data.  In contrast,
a Gaussian distribution is
{\it homoskedastic}, with a constant expectation value of sample variance.
 A `Box-Cox' plot can be used to determine the type of transformation needed
to make the data Gaussian (and hence homoskedastic).  The logarithms of
variances of segments of data are plotted against the logarithms of means, and a slope 
of 2 in the plot (equivalent to a linear scaling of rms with mean) corresponds
to a logarithmic transformation of the data \citep{box64}. Our demonstration that a linear
rms-flux relation on all time-scales corresponds to a lognormal distribution of fluxes
is effectively an extension of this result to correlated time-series data.

\subsection{Simulating non-linear light curves with an exponential model}
\label{expsims}
\begin{figure}
\begin{center}
{\epsfxsize 0.9\hsize
 \leavevmode
 \epsffile{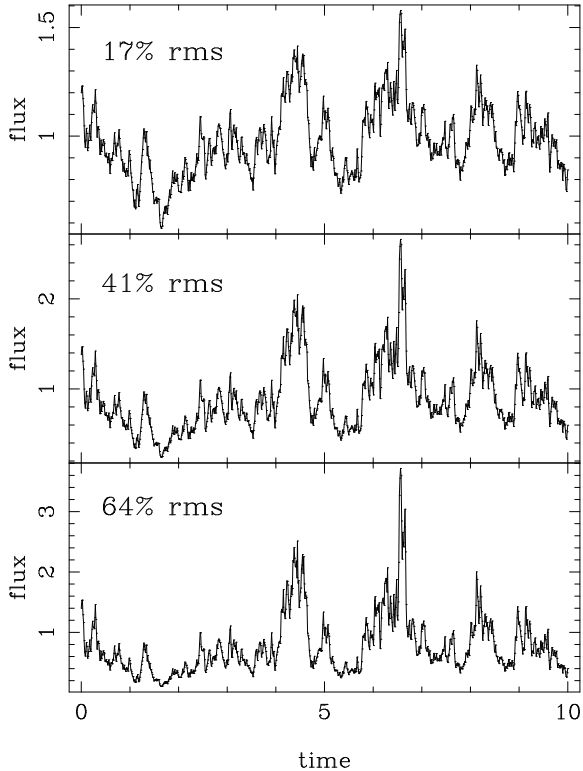}
}\caption{Simulated exponential-model light curves.  The panels show 
light curves with the same temporal structure (i.e. 
the same random number sequence is used in their generation), but 
increasing amplitude and skewness.} \label{explcs}
\end{center}
\end{figure}

\begin{figure}
\begin{center}
{\epsfxsize 0.9\hsize
 \leavevmode
 \epsffile{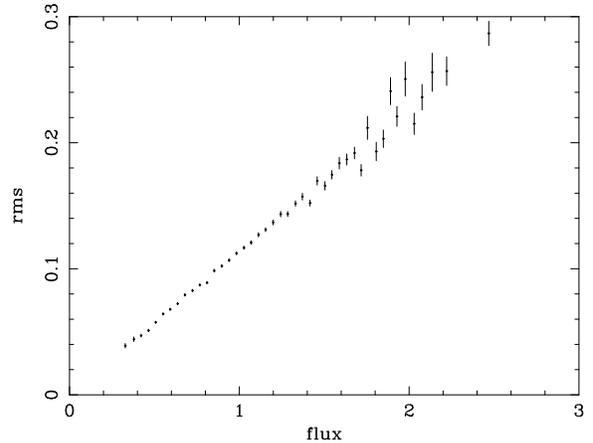}
}\caption{The rms-flux relation produced from a simulated exponential-model
light curve} \label{simrmsflux}
\end{center}
\end{figure}

We can simulate a non-linear aperiodic light curve with a linear rms-flux
relation on all time-scales by first simulating a linear aperiodic
light curve with mean 0 (e.g. using the method of \citealt{tim95}, which uses a Fast Fourier
technique based on the generalised linear model for stochastic light curves shown in 
Equation~\ref{eqn:sum})
and then calculating the exponential of the light curve at each point
in the series. 
The simple mathematical transformation to obtain exponential-model light curves
from linear
light curves implies that
the more variable (in fractional rms)
an exponential-model light curve is, the more strongly
non-linear it will appear.  This is because variations in the linear
light curve above the mean
are enhanced in the exponential-model light curve, 
compared with variations in the light curve below the
mean which are suppressed.  Thus when the variations above and below
the mean are larger, as is the case when the fractional rms is
increased, the flares in the light curve are more strongly exaggerated
compared to the dips. 
We demonstrate this property of the light curves
in Fig.~\ref{explcs}, which shows exponential-model light curves generated using
the same random number sequence (i.e. with the same `events'), but different
fractional rms.  Fig.~\ref{simrmsflux} shows the
linear rms-flux relation obtained from a simulated
exponential-model light curve like those shown in Fig.~\ref{explcs}, albeit 
of much longer duration.

The non-linearity in the light curves can also be understood in terms
of the lognormal distribution of fluxes,
because the lognormal distribution is positively skewed and
therefore can show extreme, high values of the flux which would not be
expected if the process were Gaussian.  Note however the caveat that
if the input data are not stationary (i.e. there are still trends
on the longest time-scales in the input time-series) then the
resulting distribution will not be lognormal, since the input 
Gaussian distribution is only fully sampled
on the longest time-scales, as the data become asymptotically stationary
\citep{pri82}.

It is important to note that the exponential transformation of the
input linear light curve $l(t)$ will produce a data set with different statistical
properties to the input light curve.  Therefore any attempt to simulate real
data must account for these effects, so that the correct variance and
PSD are contained in the simulated data, in order to match with the observed variance 
and PSD.  Because the output non-linear light curve
has a lognormal distribution, the basic
statistical properties (the {\it moments}) of the output can be
obtained with respect to the properties of the input linear light curve, using
the well-known results for lognormal distributions \citep{ait57,cro88}. 
For example, since in our model the input mean $\mu_{l}=0$,
the mean of the output data $\mu_{x}$ is given by:
\begin{equation}
\mu_{x}=\exp [\frac{1}{2}\sigma_{l}^{2}],
\end{equation}
where $\sigma_{l}^{2}$ is the variance of the
linear, Gaussian input data ($l(t)$).  The variance of the output data
$\sigma_{x}^{2}$ is given by:
\begin{equation}
\sigma_{x}^{2}=\exp  [\sigma_{l}^{2}]\left(\exp [\sigma_{l}^{2}] -1\right)
\label{varcorrect}
\end{equation}
and hence the fractional rms $\sigma_{\rm frac}$ of the output is simply $\sqrt{\exp
[\sigma_{l}^{2}] - 1}$.  The skewness of the distribution of
fluxes, $\gamma_{1}$, is then given by:
\begin{equation}
\gamma_{1}=\sigma_{\rm frac} \left( \sigma_{\rm frac}^{2}+3 \right).
\end{equation}
Obviously, light curves with a larger fractional rms will show more
skewed flux distributions.

It is more difficult to determine the effect of the exponential
transformation on the shape of the PSD, relative to the PSD of the input
light curve.  We demonstrate the effect on the PSD shape using simulations in
Appendix~\ref{exppsdeff}.  We note here however that 
for broad continuum type PSDs
observed in XRBs and AGN, the effect is fairly small, and the
shape of the output light curve PSD is similar to that of the input light curve.

\section{Comparison of the model with observations of Cyg~X-1}
\label{modcomp}
We now compare the predictions of our model with real data,
specifically observations of Cyg~X-1 and in particular the {\it RXTE} long-look
observation of 1996 Dec~16-19 shown in Fig.~\ref{rmslc}.  This
observation represents the longest quasi-continuous exposure (subject mainly to
orbital gaps due to Earth occultation), obtained
while all 5 Proportional Counter Units (PCUs) were switched on
(i.e. for optimum signal-to-noise).  Furthermore the PSD shape appears to be 
constant during this time, so
that the maximum amount of data can be combined without the need for
separate analysis of data with different timing properties.  Where we use
these data, we have used a 2-13~keV light curve extracted from all 5 PCUs. 

\begin{figure*}
\begin{center}
{\epsfxsize 0.7\hsize
 \leavevmode
 \epsffile{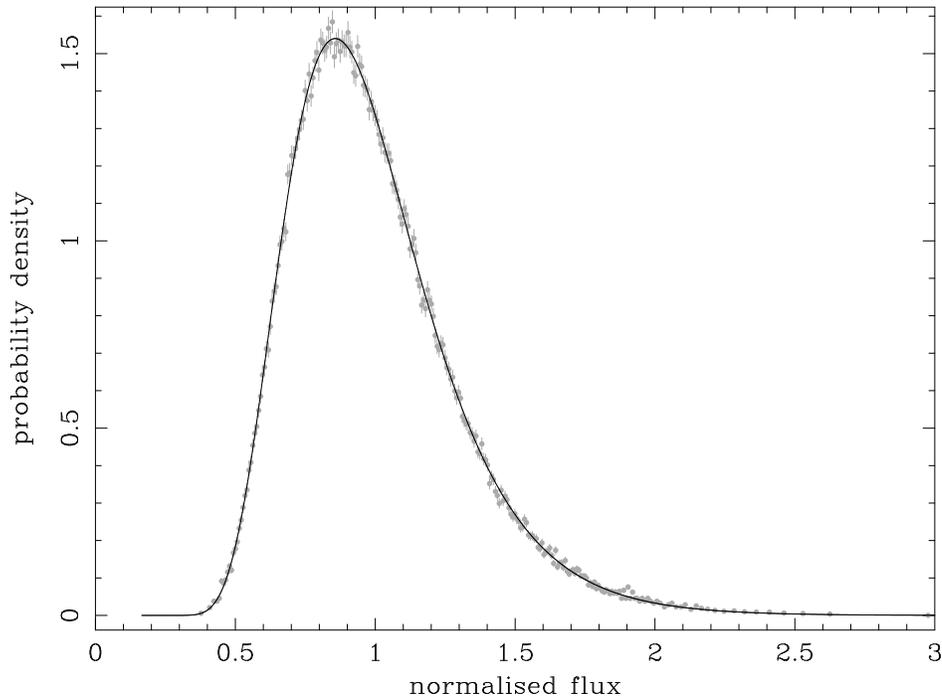}
}\caption{Flux distribution of Cyg~X-1 in 1996 Dec (grey data points), expressed as a probability
density function (PDF).
A minimum of 100 flux measurements were included in
each bin and the error bar on a bin (prior to normalisation) is then given
by $\sqrt{N}$
where $N$ is the number of fluxes in the bin.  The black line shows
the best-fitting lognormal distribution (see text for details).} \label{cygfluxdist}
\end{center}
\end{figure*}

\subsection{The lognormal distribution of fluxes}
The simplest prediction of our model is that the observed distribution
of fluxes should be lognormal.  As already noted, shot modelling of X-ray light
curves of Cyg~X-1 yielded evidence for a lognormal distribution
of shot amplitudes \citep{neg02}.  However, our model
predicts a more simple outcome that the X-ray fluxes themselves should have
a lognormal distribution, irrespective of any shot modelling of the
data.  To test this possibility, we binned up the December 1996 2-13~keV
light curve of Cyg~X-1 into 0.125~s bins, so that the typical
signal-to-noise ratio in each bin exceeds 20.  As can be seen in
Fig.~\ref{rmslc}, the light curve is not stationary even on the
relatively long time-scales observed, since clear long-term flux variations
can be seen.  To minimise the 
effects of the weak non-stationarity on the flux-distribution, we selected the
region of the light curve in the range 77-136~ks from the start
of the observation\footnote{The resulting light curve has a
mean (background-subtracted) flux of 4570~count~s$^{-1}$ and contains
253144 data points.}, which has a relatively constant time-averaged flux (on time-scales of hours),
and measured the flux distribution of the data, normalising the fluxes
by the mean.  The resulting probability density function of the
data\footnote{i.e. data points per flux bin normalised by flux bin width and
the total number of data points, so that the area under the plot is
unity} is shown in Fig.~\ref{cygfluxdist}.  Also shown is the best-fitting
three-parameter lognormal fit to the data.  The fit
is not formally acceptable ($\chi^{2}=384$ for 247
degrees of freedom), as might be expected given that the data are
weakly non-stationary, together with the additional small
Poisson component to the distribution expected from counting
statistics, which is not modelled in our fit.  
However, the fit is still remarkably good considering these
factors.  

Interestingly, the lognormal fit requires the offset parameter
$\tau$ (see Equation~\ref{lognormeqn}) to be non-zero.  In normalised flux we find $\tau=0.16$,  
consistent with the constant offset of $\sim800$~count~s$^{-1}$ observed in the
rms-flux relation for these data, and also observed in other XRBs,
which may correspond to a component with a constant flux,
or a weakly-varying component with a
constant-rms \citep{utt01,gle04}. 
The fact that the fit is remarkably close to lognormal
suggests that any {\it additive} variable component cannot vary very
strongly, otherwise it would distort the distribution significantly away
from lognormal, because lognormality is preserved multiplicatively but
not additively \citep{cro88}.  We conclude that our basic
prediction that the observed X-ray fluxes should have a lognormal
distribution is satisfied by the data.

\subsection{Powerful millisecond flares in Cyg~X-1} \label{msflares}
Recently \citet{gie03} (henceforth GZ03) reported the detection of 13 powerful
millisecond flares in 2.3~Ms of {\it Rossi X-ray Timing Explorer}
({\it RXTE}) observations
of Cyg~X-1.  The flares corresponded to rapid X-ray flux increases of a
factor of 5-10 or more on time-scales of $\sim0.1$~s during the low/hard state, and only a few
ms in the case of the most extreme event, observed during a high/soft
state.  The flares reported by GZ03 were much
larger and more common than expected, assuming
the underlying variability process is linear and Gaussian.  However,
our exponential model for the aperiodic variability in XRBs predicts
that such extreme events might occur if the fractional rms of the
source is sufficiently high.  To test this possibility, we applied the
method of GZ03 for detecting powerful flares to
simulated exponential-model light curves.  We first note that of the
12 flares detected in the low/hard state, 4 were observed in the single
long ($\sim90$~ks exposure) {\it RXTE} observation of Cyg~X-1 in
December 1996.  We therefore measured the 2-13~keV PSD of this data
set (plotted in Fig.~\ref{dec1696psd}) for use as the underlying PSD
model for light curve simulation,
to simulate exponential-model light curves to search for powerful
flares (we created a continuous PSD over the required frequency range
by interpolating between adjacent
data points and extrapolating the high-frequency PSD power-law slope measured 
between 10 and 20 Hz to higher
frequencies).  The presence of the constant (or weakly varying) component
revealed by the
lognormal fit to the distribution of fluxes (contributing 16~per~cent of
the mean flux), should dilute the
variability, so that the fractional rms of the component producing the
rms-flux relation is larger than measured
directly from the light curve.  We therefore increased the
normalisation of the input PSD by a factor $(1-0.16)^{-2}=1.42$ to
account for this effect.  We
simulated 1000 light curves
(free of Poisson noise) of length 1024~s and time resolution
2$^{-8}$~s (about 4~ms).  Following the prescription of GZ03,
we rebinned the simulated light curves into 0.125~s bins,
and measured the absolute rms, $\sigma$ of each light curve in separate
128~s segments.  We then searched each segment for bins where the flux
lies $>10\sigma$ above the mean flux of the segment.  Such extreme
events are not expected if the simulated light curve is Gaussian (i.e. without
taking the exponetial of the data).

\begin{figure}
\begin{center}
{\epsfxsize 0.9\hsize
 \leavevmode
 \epsffile{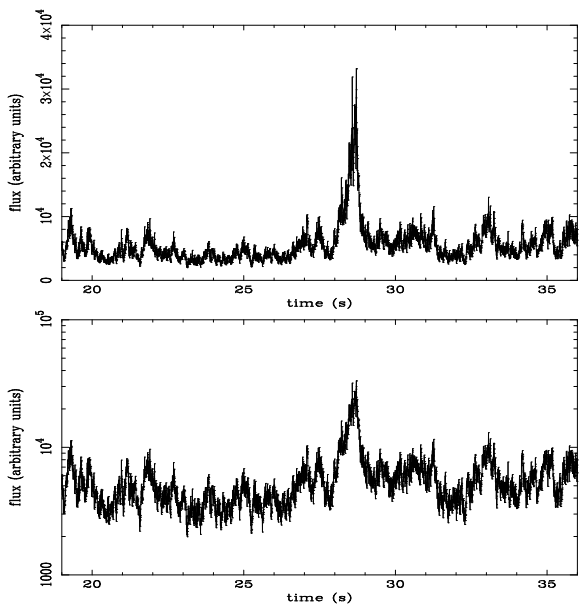}
}\caption{Powerful flare ($11.8\sigma$) in simulated `Cyg~X-1' exponential model
data (see text for details),
plotted on linear (top) and logarithmic (bottom) axes.} \label{simflare}
\end{center}
\end{figure}

From our simulation, we find 7 powerful flares in 1.024~Ms, consistent
with the 13 flares observed in $\sim2$~Ms of real data by \citet{gie03}. 
Since the flares are independent, and given the
simulated flaring rate, we can estimate that
the chance of 4 or more such flares occuring in a 90~ksec exposure (as
observed) is about
10~per~cent.  We note that the observed deviations of the
simulated flares from the mean are in the range 10.2-11.3$\sigma$,
which is roughly consistent with the low/hard state flares observed by
GZ03, which extend up to $\sigma=11.8$.
We plot light curves showing one of the
simulated flares in Fig.~\ref{simflare}, with y-axes plotted
in both linear and log units.  Note that in logarithmic units, the
light curve of the flare appears to be linear (as expected given the
exponential model used), similar to the top two panels of Fig.~1 in GZ03.

GZ03 suggest that the number of extreme flares they
observe in the low/hard state of Cyg~X-1 are consistent with the
numbers expected if the log-normal distribution of smaller shot events
proposed by Negoro \& Mineshige (2002) is extended to large amplitude
flares.  Therefore, the powerful flares observed in the low/hard state
are likely to be associated with the same variability process which
produces the smaller flares, i.e. the `normal' X-ray variability.
This picture is entirely consistent with the exponential model we
present here, which also predicts a lognormal distribution of fluxes (as 
observed in Fig.~\ref{cygfluxdist}).
Clearly, extreme events are to be expected occasionally, 
provided the average fractional rms variability is large enough.

Finally, we note that we have also attempted to replicate the very
large (12.8$\sigma$) flare observed in a high/soft state observation
of Cyg~X-1 by GZ03, using the observed PSD shape in that observation as the
underlying model to simulate light curves with a variety of values of
fractional rms.  We found that for reasonable values of fractional
rms (i.e. less than 100~per~cent), such rapid, large amplitude flares could
not be reproduced.  Therefore this particular event may have a
different origin to the normal variability in the high/soft state.

\subsection{The bicoherence of black hole XRBs}
\label{bicoh}
\citet{mac02} (henceforth MC02) have examined the higher order variability
properties of the BHXRBs Cygnus~X-1 and GX~339-4, using the
time-skewness function, which searches for time-asymmetry of the light
curve (e.g. due to different rise and decay time-scales of variations)
and also the bicoherence, which quantifies the coupling between
variations on different time-scales.  The light curves we simulate
here are time symmetric (unlike the real light curves investigated by
MC02) but presumably the time symmetry of light curves
is a function of the rise or decay time-scales of the events which
form the light curves and is not necessarily related to non-linearity
in the underlying process.  In contrast, the bicoherence is a good measure
of non-linearity in light curves, so we now investigate whether our
exponential model can reproduce the results of MC02
who found a significant coupling between variations on a broad
range of time-scales, with the strength of the coupling decreasing at
higher temporal frequencies.
\begin{figure*}
\begin{center}
{\epsfxsize 0.9\hsize
 \leavevmode
 \epsffile{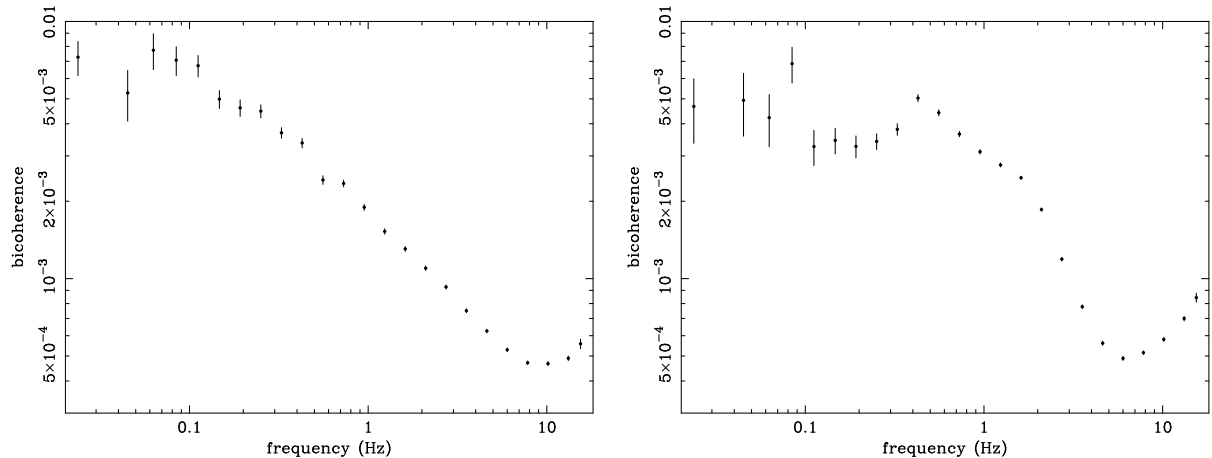}
}\caption{Left panel: bicoherence of simulated exponential model data (see text
for details).  Right panel: bicoherence of 1996 Dec 2-13~keV Cyg X-1 light
curve (see text for details).} \label{bicohsim}
\end{center}
\end{figure*}

Formally, the bicoherence measures the degree of coupling between time-series variations at
the bifrequency $(k,\:l)$, by measuring the correlation of the phase
of the signal at the frequency $k+l$, with the sum of phases of the signal at
$k$ and $l$.  Therefore the bicoherence can provide a measure of the strength of the coupling
between variations on different time-scales in the light curve (e.g. due to the signals
on different time-scales multiplying together).
We describe the calculation of bicoherence and its meaning in more
detail in Appendix~\ref{bicohapp} (also see \citealt{fac96}). 
The important thing to note here is that a non-linear
process exhibits a bicoherence which varies with the bifrequency, whereas a 
linear process exhibits a constant bicoherence (which is zero for a
linear, Gaussian process, \citealt{sub80,hin82}). 

To demonstrate the form of bicoherence predicted by our model, we
simulated a light curve of 90~ks duration with $2^{-5}$~s time
resolution, using the same PSD as the
1996 December observation of Cyg~X-1, also used in the simulations in
Section~\ref{msflares}.  As with the simulations in
Section~\ref{msflares}, the PSD normalisation of the
simulated variable light curve was corrected to take account of the 
possible constant
component in the observed
light curve and the constant component added at the end of
the simulation to obtain the correct flux level before Poisson noise
was added (to account for the additional bicoherence produced by
counting statistics, see Appendix~\ref{bicohapp}).  We measured the
bicoherence using segments of 256~s duration and corrected for noise effects
and bias using the prescription given in Appendix~\ref{bicohapp}. 
In order to reduce the noise and present the bicoherence as a
simple 2-dimensional plot, we followed the approach of
MC02 and averaged the bicoherence measurements
according to the sum of the frequencies $k$, $l$.  We then further binned
the bicoherence into logarithmically spaced frequency bins 
(factor of 1.3 spacing between
bins) so that there is a minimum of 20 individual bicoherence
measurements (i.e. before averaging according to $k+l$) per bin.
We calculated standard errors using the spread in data in each bin.

The resulting plot of binned bicoherence is shown in
Fig.~\ref{bicohsim} (left panel).  For comparison we also show the bicoherence of
the entire 1996 Dec 2-13~keV light curve of Cyg~X-1
(Fig.~\ref{bicohsim} right panel, c.f. Fig.~3 of
MC02 for a similar plot).  Clearly
the most general features of the observed bicoherence, such as a
relatively flat shape at low frequencies and a reduction
in bicoherence at high frequencies, are well reproduced by our model.
This is to be expected, since the multiplication of sines which our
model is derived from implies that the strength of coupling between
two signals is proportional to the product of the power in the two
signals.  Thus the bicoherence is flat below $\sim0.1$~Hz because the
power is constant below that frequency, and the bicoherence decreases
at higher frequencies because the power decreases towards higher
frequencies.  Note the increase in bicoherence above
$\sim10$~Hz in both plots, which is a result of the photon
counting statistics (see Appendix~\ref{bicohapp}).

The normalisation of the bicoherence produced by our
model is of a similar magnitude to that observed in real data.  The
value of the normalisation is clearly a function of the power
which is coupled together, hence in our model a larger fractional rms will
produce a larger bicoherence.  We
note that in the case of a broadband aperiodic signal such as
measured here, the normalisation is not only a function of the coupling
strength, but also a function of the duration of segments used to
calculate the bicoherence (see \citealt{gre88} for a detailed discussion
of this point).   This effect is caused by
the spreading out of signal power which contributes to the
bicoherence.  As the duration of a segment is increased, so the number
of Fourier frequencies used to calculate bicoherence also increases
and (in the case of a broadband aperiodic signal) although the power-density
at each Fourier frequency does not change systematically,
the signal power at each
frequency decreases accordingly.

Although the most general characteristics of the bicoherence predicted by
our model are similar to those of the observed bicoherence, there are
substantial differences in detail.  In particular, the observed
bicoherence only decreases substantially above a few Hz, not $\sim0.1$~Hz as in our
model.  Also, clear bumps can be seen in the observed bicoherence,
which may be a result of QPOs which are also coupled together
(although it is not yet clear why the bump at $\sim0.4$~Hz is offset
from the similar feature in the PSD at $\sim0.8$~Hz ).  Therefore, 
although our model can
reproduce the general properties of the observed variability such as its
lognormal and non-linear behaviour, that is not the whole story.  The
significant differences between the bicoherence predicted by our model
and the observed bicoherence imply that there may be more complex
interactions occuring between different variability time-scales than
the simple multiplicative
coupling assumed by our exponential model.  For example, since
the bicoherence essentially measures correlations between phases at different
temporal frequencies, a number of mechanisms can produce significant values of
bicoherence without significantly distorting
the flux distribution away from lognormal, e.g. frequency modulation 
\citep{ria00} or phase locking of two signals.  Such effects could 
explain the deviation of the real
bicoherence spectrum away from that predicted by our model, while
maintaining the observed lognormal distribution and linear rms-flux relation.

\section{Discussion}
\label{discuss}
We have shown that the linear rms-flux relation observed in the light curves of XRBs and AGN, if it applies on 
all time-scales, implies that the light curves ($x(t)$) are formally non-linear, being described
by the simple transformation $x(t)=\exp[l(t)]$, where $l(t)$ is a linear, Gaussian time series.  The resulting flux
distribution for stationary data is lognormal, and we have confirmed 
this prediction using data from an X-ray light curve of Cyg~X-1.  We have also shown that the
powerful flares observed in the low/hard state \citep{gie03}, occur naturally in our
phenomenological model for the variability.  And we have shown that the amplitude and general shape
of the bicoherence function \citep{mac02} can be explained by our model.  Thus many of the observed 
non-linear properties of the light curves appear to be due to the same underlying process. 
In this Section, we will 
discuss the implications of these results for AGN (which we have only briefly touched on so far),
consider constraints on the physical nature of the variability process,
and finally discuss implications of these results for future modelling of the variability.

\subsection{Non-linearity in AGN variability: NLS1 and low states in NGC~4051}
In Section~\ref{modcomp}, we showed
that our exponential model can explain some of the characteristics of
variability in Cyg~X-1 (and likely also other XRBs).  It is also natural to consider the implications
of this phenomenological model for the interpretation of AGN variability, since AGN also
show rms-flux relations in their light curves, and similar PSDs to
X-ray binaries (e.g. \citealt{vau03a,mch04}).   
The exponential model of X-ray variability provides a natural
phenomenological explanation for the
X-ray light curves of certain NLS1, which appear to be more strongly non-linear
than those of less variable AGN (e.g. \citealt{bol97,bra99,lei99,dew02}).  In fact,
\citet{gas03} has demonstrated that the X-ray fluxes in the NLS1 IRAS~13224-3809 are well described
by a lognormal distribution, as expected if our exponential model also applies to NLS1 X-ray
variability.  Figure~\ref{explcs}
shows how exponential-model light curves with greater fractional rms
variability (as typically observed in NLS1) 
naturally appear more non-linear than those with lower
fractional rms, even though the same model is used to generate
all the light curves.  In fact, it is interesting to note the general
similarity of the most variable exponentiated light curve in
Fig.~\ref{explcs} to some observed light curves of NLS~1,
e.g. that of IRAS~13224-3809 \citep{dew02,gas03} and also NGC~4051
\citep{mch04}. 

Therefore, our model suggests that
non-linear X-ray variability is common to most (if not all) AGN but is
only readily observed in NLS1 due to their enhanced fractional rms. 
This scenario suggests that variability processes in NLS1 are similar to
those seen in other Seyferts, with the major difference being their
amplitudes of variability, rather than the physical nature of the variability
itself.  It is interesting to note that the amplitude of variations in some NLS1 
is particularly large
(e.g. \citealt{bra99}),
large enough to violate the well known constraints on rapid variability imposed by 
radiative efficiency arguments \citep{fab79}.  \citet{bra99} note a number of ways in which these 
constraints can be circumvented (e.g. if radiation is released from multiple locations, which may 
be possible if the X-ray variability is externally triggered, perhaps by fluctuations in the accretion flow).  
However, if we assume that these extreme variations are the result of the same non-linear process that
operates in less variable objects, or when the source is more `quiet' (i.e. the extreme events are the high-flux 
tail of a highly skewed lognormal distribution), it remains an open question as to how extreme these 
events can be.  Lognormal distributions are mathematical entities, but what we observe must
be constrained in some way by source physics - does the physics of the source intervene to 
prevent extreme tails in lognormal distributions?
Long monitoring observations of these AGN, which would sample
a wide range of fluxes, can help to answer this question.  A further related issue is
whether such extreme events can occur frequently in some XRBs (as opposed to the very rare events 
seen in Cyg~X-1, GZ03).  On the equivalent 
time-scales (i.e. less than seconds, if we scale with black hole mass) XRBs do not show
high enough rms to appear similar to light curves of the more extreme
NLS1.  We have demonstrated that the extreme
non-linearity seen in NLS1 light curves probably results from their large variability amplitudes,
which imply a more skewed distribution of fluxes.  The question then is not `why is NLS1 X-ray variability
non-linear' but rather `why are NLS1 so variable in the first place?'

One characteristic feature of aperiodic `red-noise' light curves is their
self-similar (technically, {\it self-affine}) nature over a broad range of time-scales.  
For example, in the particular case of `flickering'
where the PSD has an index of -1 over a broad range of time-scales,
the variability observed on long time-scales
(e.g. years) looks very much like the variability observed on much
shorter time-scales, e.g. days (the light curves are scale-invariant).  This property of
flickering light
curves means that the exponential model light curves shown in
Fig.~\ref{explcs} might just as well represent the long-term (years) X-ray light
curves of AGN as they do the shorter-term light curves, provided the AGN
show significant variability power on long time-scales.  
In this case, the prolonged low-flux, low-variability periods observed when the fractional rms is
high would appear similar to the prolonged (weeks to months) low flux states observed in
the NLS~1 NGC~4051 \citep{utt99,utt03,utt04b}.  In fact, it is interesting to note
the similarity between the long-term light curve of NGC~4051 and the
light curve on much shorter time-scales (e.g. see \citealt{mch04}), which
further indicates the scale-invariance of the light curves.  

If the low states in NGC~4051 are simply the continuation of the non-linear form of AGN
light curves to long time-scales, then they represent another
manifestation of the scale-invariance of flickering light curves and
may not be physically distinct states after all, i.e. these states are
not distinct in the sense that the low/hard and high/soft states observed
in BHXRBs are (see also \citealt{utt03,utt04b} for spectral evidence to support
this conjecture).  \citet{gas03} has also raised a similar point regarding the low-variability,
low-flux periods observed on shorter time-scales (days) in the light curve of IRAS~13224-3809.
Regardless of the physical interpretation of the variability, our
phenomenological model implies that low states such as those observed in
NGC~4051 may be common in other AGN with large variability amplitudes,
if the variability also extends to long time-scales.

\subsection{Implications for physical models}
\label{varmod}
Our original suggestion that the rms-flux relation rules out additive shot-noise models \citep{utt01}
is confirmed by the analysis presented here, since additive shot-noise models are inherently linear and
cannot reproduce the 
non-linearity and lognormal flux distributions implied by the rms-flux relation. 
Furthermore, the
fact that the flux distribution is lognormal places fairly stringent constraints on any physical models
for the variability, since it implies that the underlying physical process is multiplicative, rather
than additive.  If the total X-ray emission were produced by many independent elements adding together
(e.g. from separate active emitting regions in the corona), then, subject to certain caveats,
irrespective of the distribution of flux from each element the resulting flux distribution would 
tend towards a Gaussian distribution (due to the central limit theorem).  The main caveat to this statement is that
the distribution of fluxes from each independent element is not so skewed that only a few elements
contribute significantly to the flux at any one time, e.g. if the distribution of
flux from each additive element is itself lognormal and highly skewed.  

To test the effect
on the observed distribution of adding together fluxes from multiple, independent
elements with lognormal flux distributions, we simulated light curves with the same
length and time resolution as the quasi-stationary light curve used to make the 
observed flux distribution in Fig.~\ref{cygfluxdist}, but made from a sum of
exponential model light curves with PSD shape identical to that observed (see section~\ref{modcomp}). 
Since the
total variance is the sum of the individual light curve variances, for $N$ elements with equal 
mean fluxes and fractional rms, the fractional
rms of each element is equal to the total fractional rms multiplied by $\sqrt{N}$. 
Thus to produce a given observed fractional rms,
the flux distribution from individual elements becomes more skewed as the number of elements increases
so that only a few elements can contribute significantly to the total flux, even for large $N$.

\begin{figure*}
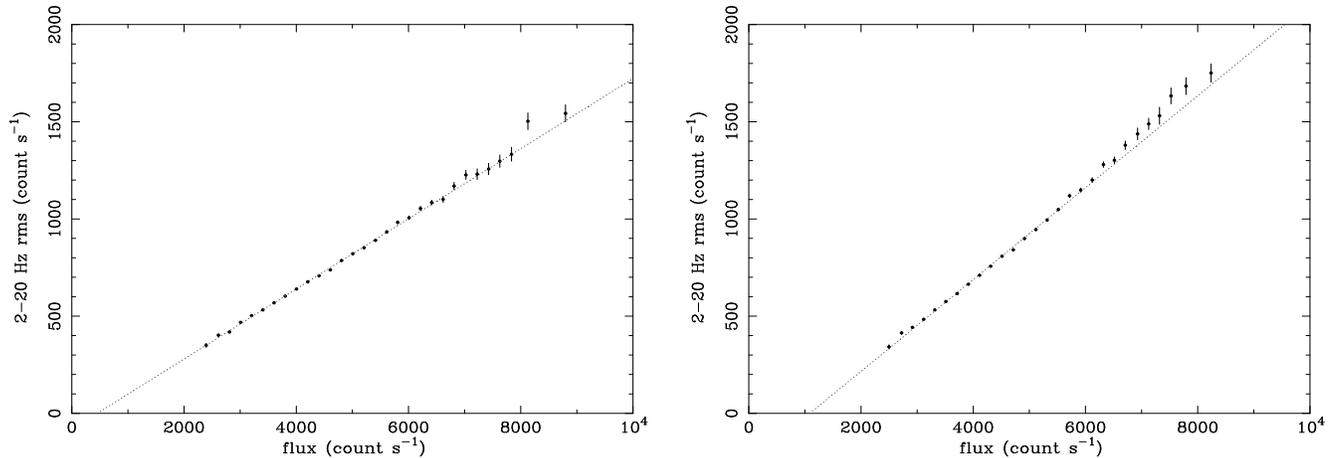

\begin{center}
%\vspace{-0.5cm}
\hbox{
\psfig{figure=cygx1rmsobs.ps,width=8.5cm,height=6cm,angle=-90}
\hspace{0.3cm}
\psfig{figure=cygx1rms5compsim.ps,width=8.5cm,height=6cm,angle=-90}
}
\vspace{-0.5cm}
\end{center}
\caption{Comparison of rms-flux relations for real Cyg~X-1 data (left) and simulated data (right) consisting
of 5 independently varying light curve elements with equal mean flux and fractional rms.  The grey dotted
lines denote the best-fitting linear models (see text for details).}
\label{multcomprms}
\vspace{-0.3cm}
\end{figure*}

Our simulations show that for $N\leq5$, the resulting total flux distributions show reduced chi-squared
$\chi^{2}_{\nu}<1.5$, i.e. comparable to that observed in the real data.  We note however that
the rms-flux relations of our simulated light curves show significant deviations from linearity
for $N>2$.  For example, in Fig.~\ref{multcomprms} we show the rms-flux relations
for the observed 1996~Dec quasi-stationary data and our simulation for $N=5$.  The
simulated rms-flux relation shows a significant systematic deviation from a linear relation at higher
fluxes which is not observed in the real data.

The deviation from a linear relation is caused
by the fact that the flux distributions of the individual elements are highly skewed. At low fluxes, 
all the elements contribute significantly to the light curve, and the fractional rms
is diluted accordingly (since the elements vary independently).  However, at high
fluxes only a small number of components dominate the variability and hence 
the fractional rms (equivalent to the gradient of the rms-flux relation) increases.
 A linear fit to the simulated rms-flux relation
gives reduced chi-squared of $\chi^{2}_{\nu}=3.5$ for 26 degrees of freedom, compared with 
$\chi^{2}_{\nu}=1.5$ for 28 degrees of freedom for the real data.  Only simulations with $N=2$ give fits
as good as those observed, so we conclude that the number of independently varying elements is
restricted by the linearity of the rms-flux relation to be less than three.
The simplest explanation of the data is that the X-ray emission primarily originates 
from a single coherent emitting region, with flux from different parts
of the emitting region being modulated in
the same way (i.e. there is little contribution from
independent flares), or from multiple regions that are
somehow connected, so they do not vary independently (e.g. if separate regions are driven by the same
underlying variations in the accretion flow).  

The observed lognormal flux distribution also rules out SOC models for variability, 
which predict a power law distribution of individual
shot {\it fluences} \citep{bak88,chr91}.  The resulting
distribution of the total fluxes is therefore a sum of elements with 
power-law distributions, with the number of summed elements depending on how much
the individual shots overlap one another \citep{chr92}.  The expected flux distribution is therefore
intermediate in shape between
a power-law and Gaussian distribution, unlike the lognormal distribution observed here.
Using a `shot-fitting' approach
to quantify variability, \citet{tak95} have demonstrated that the observed distribution of shot fluences
in Cyg~X-1 is better fitted by an exponential rather than the power-law distribution expected
from SOC models.  This result is probably related to the fact that the flux distribution is lognormal, and that the
tail of the lognormal distribution (which is dominated by large `events' which are easily picked out by
shot-fitting methods) approximates an exponential distribution (see also discussion in \citealt{neg02}).
To account for this different distribution (and also to produce more realistic PSD indices), \citet{tak95} modify 
the SOC model to incorporate gradual diffusion of matter through the disk.  The resulting model bears
some resemblance to models where the variability arises from stochastic variations in the accretion flow,
rather than from the deterministic (if unpredictable)
dynamics of any critical state\footnote{As \citet{tak95} note, in a strict sense the resulting
behaviour is no longer SOC, because
a key characteristic of the SOC state is a power-law distribution of shot fluences.}.  We also
note here the temporal variability component of the
`thundercloud model' of \citet{mer01} which, although an additive shot-noise model,
successfully produces a linear
rms-flux relation.  However, the thundercloud model
assumes a power-law distribution of shot sizes (i.e. similar to the 
distribution produced by SOC models), and hence should be ruled out for the same reason as SOC models,
i.e. the flux distribution is not lognormal.
The reason why models with skewed (but not lognormal) flux distributions can produce linear
rms-flux relations is discussed further in Appendix~\ref{rmsdistrib}.

More generally, the fact that the data are consistent with a {\it static} non-linear transform of stochastic,
linear input data suggests that deterministic, non-linear types of model
such as dynamical chaos are not required to explain the data.  For example,
a possible interpretation of apparent non-linear behaviour in light curves is that it
is a result of unpredictable chaotic behaviour (the non-linearity is said to be
{\it dynamical}).  Such behaviour might be the signature of a rather simple physical system,
which can be
 described by simple dynamical equations.  In fact, searches for chaotic dynamics
(i.e. a `low-dimensional attractor') in Cyg~X-1 X-ray light curves found no evidence for 
such behaviour \citep{loc89}.  On the other hand, possible evidence for chaotic behaviour
has been suggested for the X-ray lightcurves of the microquasar
GRS~1915+105 \citep{mis04}.  However,
\citet{the92} note that a succession of nested hypotheses should be tested before
evidence for chaos is assumed, including whether or not the variability is consistent with
static non-linearity (see also \citealt{kan97}).  Therefore, it is possible that the static
non-linear behaviour
which we have demonstrated in this paper could be mistaken for dynamical chaos.

Having ruled out a large number of models, we are left wondering which models are still
permitted by the data.  As noted by \citet{gas03} (and see also \citealt{cro88}), lognormal
distributions are very common in Nature, because they can be produced in a number of different ways
(all involving multiplicative processes).  For example, comminutive processes, involving
the random splitting apart or fracturing of objects (e.g. the crushing of rocks) lead to a lognormal
distribution of sizes, because the probability of a given size is dependent on a multiplicative
sequence of fracture events.  One could similarly envisage how multiple emitting regions
following a lognormal distribution
of sizes (or equivalently, fluxes) might be produced.  However, it is difficult to see how
such an arrangement of emitting region sizes could lead to a lognormal distribution of the total flux,
unless we somehow only witness one of the emitting regions at any given time (otherwise
the observed distribution would be a sum of lognormals, a possibility we have ruled out earlier
in this Section).  Another type of multiplicative model, involving the cumulative build up and release of 
energy from a reservoir, where the amount of energy released scales with total energy in the reservoir,
is suggested by the jet-disc coupling model of \citet{mal04}.  Further investigation of these
models is needed, but we note here that the lognormal form of observed data might be quite constraining for
the number of components (e.g. jet, disk) which contribute (either adding or subtracting) independently to
the reservoir.

Since the variability can be thought of as a static exponential transformation of linear, Gaussian data,
we might first consider a direct physical interpretation of this fact.  For example,
the underlying variable process (e.g. variations in accretion rate) might be linear and Gaussian, but the
X-ray emission process may be non-linear such that the observed X-ray 
variability is transformed to be non-linear
and with a lognormal distribution.  This possibility seems unlikely however, because (as we noted earlier), 
the fact that
linear rms-flux relations are seen in both black hole and neutron star XRBs, which have different
X-ray spectra and different emission mechanisms (e.g. \citealt{don03,pou03,gil03} and see
also \citealt{utt04}), makes it appear unlikely that
the emission process itself is the origin of the non-linearity and rms-flux relation.

An intuitive and simple way to produce the observed lognormal distribution in the underlying process
is suggested by the
rms-flux relation and the derivation of our exponential model which follows from it.
The amplitude modulation effect implied by the rms-flux relation suggests that variations
on different time-scales must multiply, rather than add, together.  An obvious mechanism
for coupling together variations on different time-scales is if the variations are produced
at different radii in the accretion flow, with slower variations produced at larger radii. If the 
accretion variations can then propagate to small radii (where the X-ray emitting region exists),
then variations on different time-scales can couple together, because each annulus
in the accretion flow will see variations on longer time-scales produced at larger radii.
Such a model was proposed by \citet{lyu97}, in order to explain the fact that observed
X-ray variability time-scales in XRBs extend to much longer time-scales 
than the longest time-scales expected in the inner, X-ray emitting region 
(see also \citealt{kot01,chu01,kin04} for enhancements to this model).  \citet{utt01}
then noted that the same model could explain the observed rms-flux relation - 
a situation that has since been reinforced by the discovery that the rms-flux relation
in the X-ray millisecond pulsar SAX~J1808.4-3658 is most likely produced in the 
accretion flow and not in a flaring corona \citep{utt04}. 

If the variations at a given radius are constant in fractional rms (e.g. as might be expected
if they correspond to variations in the viscosity which are independent of mass accretion rate)
then the situation is almost directly analogous to the sine multiplication picture used to derive
the exponential model in this paper (here the sines represent the variations produced at 
different disk radii, although in reality the variations are unlikely to be periodic).  Thus
the rms-flux relation, lognormal flux distribution and non-linearity in the light curves
can be explained in terms of a rather simple physical picture, which it should be noted
can also explain the observed spectral-timing properties 
(energy-dependent phase lags, PSD shape and coherence) of XRBs and AGN 
\citep{kot01,vau03a,mch04}.

\subsection{Testing future variability models}
We first stress that the X-ray emission process (e.g. Comptonisation in 
a corona) and the fundamental origin of the variability are not necessarily implicitly related.
In fact, they appear to be largely unrelated, as revealed by the fact that linear rms-flux relations are
seen in both neutron star and black hole XRBs which have different X-ray emission mechanisms 
\citep{utt04}. Therefore, in this Section, we will consider tests of models for the underlying variability
process and not explanations of spectral variability, which are beyond the scope of this work.
At the beginning of this paper, we noted that models for variability tend to start by explaining the
shape of the PSD.  We pointed out that the PSD is perhaps not the best tool to use to distinguish 
variability models, because many models can produce the required PSD shapes (e.g. using
broad distributions of shot or event time-scales) and also because a standard PSD shape is in
fact not a fundamental characteristic of real X-ray variability - XRBs, and possibly AGN
show a variety of PSD shapes and
PSD shape varies within the same state and between states.  We have demonstrated in this paper
three closely related characteristics of XRB and AGN variability that do have strong discriminating power
between models: the rms-flux relation, non-linear variability and a lognormal distribution of fluxes.
These characteristics rule out a whole swathe of models, from additive shot noise, to SOC, to
any models which consist of multiple, independent varying regions whose variations add together
to produce the observed variability.  The variability process should be multiplicative
in order to produce the observed characteristics.  

Therefore, an important test of any variability model is whether it can
reproduce the variability properties outlined in this paper.  This fact is regardless
of the specific PSD shape predicted by the model, or other properties of the variability
such as spectral-timing behaviour, which can be thought of as higher-order
features of any model.  For example, different PSD shapes can be produced by various
`filters' which act on the underlying variability process (since the observed X-ray variability
is only really a proxy for the underlying process).  E.g. if the variability is caused by accretion rate
variations, an extended distribution of the X-ray emission can act as a low-pass filter, producing
a steepening of the PSD of the underlying process at high temporal frequencies along with 
frequency-dependent lags between different energy bands
(e.g. see \citealt{kot01,zyc03,vau03a} and \citealt{mch04} for discussion).  However, the
observed non-linearity, rms-flux relation and flux distribution are characteristics
of the underlying variability process which simple filtering cannot reproduce (since the filtering
is a linear transformation of the data).

We therefore outline the following battery of tests for models of the underlying variability process.  
Beginning with the most important, these are:
\begin{enumerate}
\renewcommand{\labelenumi}{\arabic{enumi}}
\item. Do the model data show a linear rms-flux relation?
\item. Does the rms-flux relation occur on all time-scales,
or equivalently, if the model data are stationary do the model data show a lognormal
flux distribution?
\item. Does the PSD of model data match with observations?
\end{enumerate}
The first two properties to be tested are likely to be most closely associated with the underlying
variability process.  The final test, of PSD shape, which is normally applied to variability models 
(e.g. \citealt{pou99,min94}) is less fundamental in our opinion, because the precise PSD shape is
not even a unique characteristic of variability in a single source, since PSD shape varies between states 
and between observations of the same state (e.g. \citealt{pot03}).  However, the general
characteristics of the PSD shape (e.g. parameterisation of the low/hard state PSD as multiple Lorentzians
in both neutron star and black hole XRBs,
with similar correlations between characteristic frequencies \citep{bel02}) are common enough to a 
variety of sources that the PSD shape, in combination
with the first two tests, remains a useful test of the underlying variability process.

The first test, for a linear rms-flux relation, is simple to perform as an initial
check of the model, but not sufficient to determine if the non-linear properties
of the model are similar to those of real data.  It is interesting to note here
that a nearly linear rms-flux relation may be obtained from a variety of 
positively skewed flux distributions, if the amplitude of variability
in the frequency range used to measure rms is large.  We investigate the rms-flux relation
for a variety of flux distributions and PSD shapes in Appendix~\ref{rmsdistrib}.
However, if measured fractional rms is small (e.g. as found for the 2-20~Hz range
used to measure the rms-flux relation for Cyg~X-1, or the equivalent high frequencies
used to measure rms-flux relations in AGN), the shape of the
rms-flux relation is a strong function of the flux distribution, and in this case
a linear rms-flux relation is a good predictor of an underlying lognormal distribution
of fluxes.  As an additional check, we then suggest carrying out the second test.

The two forms of the second test
result from the fact that the lognormal flux distribution is
 a corollary of the fact that the rms-flux relation seems to apply on all time-scales (as is implicit in
the derivation of the exponential model).  If that were not the case, one would not observe a lognormal flux
distribution in stationary data.  However, the condition of stationarity is essential for testing for lognormality.
If the simulated data are not stationary, one must instead test different time-scales of variability
to see if the rms-flux relation applies (e.g. see methods in \citealt{utt01,gle04} and in 
Section~\ref{rmstimescale}).  For example, magnetohydrodynamic (MHD) simulations
of turbulent accretion flows are reaching the stage where they can demonstrate the 
variability expected from such flows in a physically self-consistent way (e.g. \citealt{mac03,arm03}).  
However, due to current computational constraints the simulated data sets are not long enough to 
probe the longest
variability time-scales, and hence are not stationary.  Nonetheless, it would be simple to test these
data sets to see if linear rms-flux relations are present on different time-scales.  

Finally, we note that other tests of non-linearity, such as the bicoherence, may also impose strong 
constraints on
models for the variability.  In Section~\ref{bicoh} we showed how our exponential model can explain
the amplitude and general shape of the bicoherence of Cyg~X-1, but not the detailed characteristics of the 
bicoherence.  For example, the presence of apparent bumps in the bicoherence spectrum indicates stronger 
coupling between variations on certain time-scales than we might otherwise expect given our simple model.
It is important to develop these methods of non-linear analysis to shed light on these issues, and provide
further clues for the development of future models.

\section{Conclusions}
We have expanded on the discovery of a linear relation between rms variability and flux in XRBs and AGN,
to demonstrate the following:
\begin{enumerate}
\renewcommand{\labelenumi}{\arabic{enumi}}
\item. If the linear rms-flux relation observed in XRBs and AGN applies on
all time-scales, light curves $x(t)$ are produced which show
a simple, formally
non-linear type of variability, $x(t)=\exp[l(t)]$, where $l(t)$ is a linear input light curve.
\item. Our phenomenological, `exponential model' for X-ray variability predicts a lognormal flux
distribution for stationary light curves, which we confirm using data for Cyg~X-1.
\item. The powerful millisecond flares observed in Cyg~X-1 in the low/hard state \citep{gie03}
are a natural consequence of the non-linear variability predicted by our model.
\item. Our model can reproduce the general shape and amplitude of the bicoherence spectrum 
observed in Cyg~X-1 \citep{mac02}, although detailed features in the bicoherence cannot be
explained, implying a stronger coupling between certain time-scales than we naively expect from
our model.
\item. Our model suggests that the clear non-linear X-ray variability observed in some NLS1 AGN
results from the same variability process that applies in less variable AGN (the difference is simply that NLS1 are
more variable so the non-linearity implied by our model is easier to detect).  The low flux states observed in
the NLS1 NGC~4051 are also a manifestation of the same variability process.
\item. Physical models for the variability must involve multiplicative processes, such as the
varying-accretion model of \citet{lyu97} and cannot be due to additive processes 
such as shot noise, or SOC processes, or multiple, independently-varying X-ray emitting regions.
\item. Future physical models for the variability should first be tested for the existence of
a linear rms-flux relation.  Other common tests such as PSD shape or spectral timing properties are
more likely secondary characteristics of the variability process; but the rms-flux relation and the
resulting non-linear variability and lognormal flux distribution (in stationary data), appear to be more 
fundamental features of the underlying process.
\end{enumerate}

\subsection*{Acknowledgments}
We would like to thank Tom Maccarone for useful discussions about bicoherence, and
the anonymous referee for an excellent and thorough report which improved
the content of this paper.  PU acknowledges support from PPARC
and current support from the US National Research Council.
IM$^{\rm c}$H acknowledges the support of
a PPARC Senior Research Fellowship.  SV acknowledges support from PPARC.
This research has made use of data obtained from the High Energy
Astrophysics Science Archive Research Center (HEASARC), provided by
NASA's Goddard Space Flight Center.

\appendix

\section[]{A formal proof of non-linearity}
\label{nonlinproof}
We can express the exponential model for the light curve $x(t)$
in terms of the series expansion of $\exp[l(t)]$, where $l(t)$ is the linear, 
input light curve:
\begin{equation}
x(t)=\exp[l(t)]=1+l(t)+\frac{[l(t)]^{2}}{2}+\sum_{n=3}^{\infty}\frac{[l(t)]^{n}}{n!}
\end{equation}
Next, we can change to discrete time steps and
replace each of the $l(t)$ terms using the definition of
a linear light curve in Equation~\ref{lindef}:
\begin{eqnarray}
X_{i}& = & 1+\sum^{\infty}_{j=0} g_{j}u_{i-j} + 
\frac{1}{2} \sum^{\infty}_{j=0} g_{j}u_{i-j} 
\left( \sum^{\infty}_{k=0} g_{k}u_{k-j}\right) \nonumber \\
& & + \sum_{n=3}^{\infty}\frac{\left(\sum^{\infty}_{j=0} g_{j}u_{i-j}\right)^{n}}{n!}
\end{eqnarray}
which can be re-expressed as:
\begin{eqnarray}
\label{voltnew}
X_{i} & = & 1+\sum^{\infty}_{j=0} G_{j}u_{i-j}
+\sum^{\infty}_{j=0}\sum^{\infty}_{k=0}G_{jk}u_{i-j}u_{i-k} +
\nonumber \\
& & \sum^{\infty}_{j=0}\sum^{\infty}_{k=0}\sum^{\infty}_{l=0}
G_{jkl}u_{i-j}u_{i-k}u_{i-l}+ ...
\end{eqnarray}
Where the coefficients $G_{j}, G_{jk}, G_{jkl}\ldots \propto g_{j}, g_{j}g_{k},
g_{j}g_{k}g_{l}\ldots$ and the higher-order co-efficients are non-zero. 
Equation~\ref{voltnew} is a Volterra series 
(c.f. Equation~\ref{volterra}),
an example of a formally-defined non-linear time-series model.

\section[]{The effect of the exponential transformation on PSD shape}
\label{exppsdeff}
The shape of the PSD of a linear input light curve is not preserved by
the exponential transformation envisaged by our model.  This can be understood
in terms of the series expansion of the exponential model shown in
Appendix~\ref{nonlinproof}, where the model can be expressed in terms of
the linear light curve $l(t)$ {\it plus} higher order polynomials of $l(t)$.
Multiplying signals in the time domain is equivalent to
convolving their Fourier transforms in the frequency domain, and the
effect of convolution is to transfer power to other frequencies, thus
distorting the PSD shape from its original form.  Thus the effect of the 
exponential transformation of $l(t)$ is to add power to the PSD {\it and}
transfer this power to different frequencies.
It is important to take this effect into account when simulating time
series using the exponential transformation,
if it is necessary that the simulated light curve PSD must
match some observed PSD.

\begin{figure}
\begin{center}
{\epsfxsize 0.9\hsize
 \leavevmode
 \epsffile{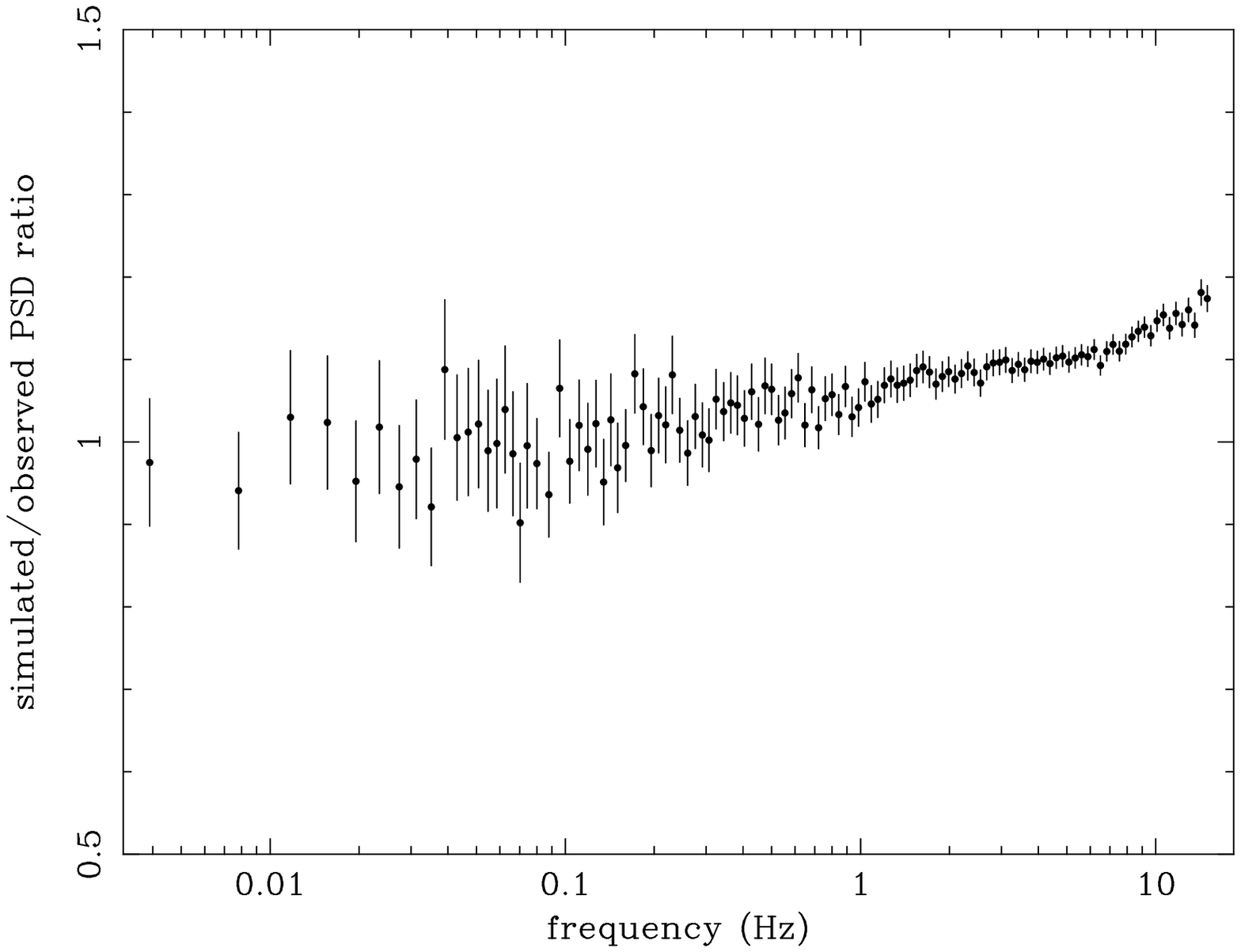}
}\caption{Ratio of the PSD of simulated light curve to observed input
PSD (prior to normalisation correction), for the broadband PSD shape plotted
in Figure~\ref{dec1696psd}.  See text for details.} \label{psdratio}
\end{center}
\end{figure}

Fortunately, for typically observed light curves, with broad continuum
PSD shapes, the distorting effect of the exponential transformation is
relatively small, so that it is reasonable to use the observed PSD as
the PSD of the input linear light curve.  In this case it is still
important to apply the appropriate
correction to the input PSD normalisation in order to return the correct
variance in the output light curve, i.e. use Equation~\ref{varcorrect}
to determine the input linear light curve variance needed to produce
the required output variance and then multiply the input PSD normalisation by
the ratio of the input variance to the observed variance. 

As the fractional rms increases, so does the distorting effect on the
PSD.  However, for most observed fractional rms (e.g. 20-40~per~cent) the
distortion is not serious.  In Fig.~\ref{psdratio} we plot the ratio
of the PSD of the simulated light curves used in Section~\ref{modcomp} 
to the PSD of the observed light curve (shown in
Figure~\ref{dec1696psd}).  The variance of the simulated light curves
(prior to dilution with the constant component) is $\sim40$~per~cent. 
The ratio is close to 1 in most cases, but increases towards high
frequencies, however as the power is small at high frequencies anyway, we
find this small amount of distortion acceptable.

\begin{figure}
\begin{center}
{\epsfxsize 0.9\hsize
 \leavevmode
 \epsffile{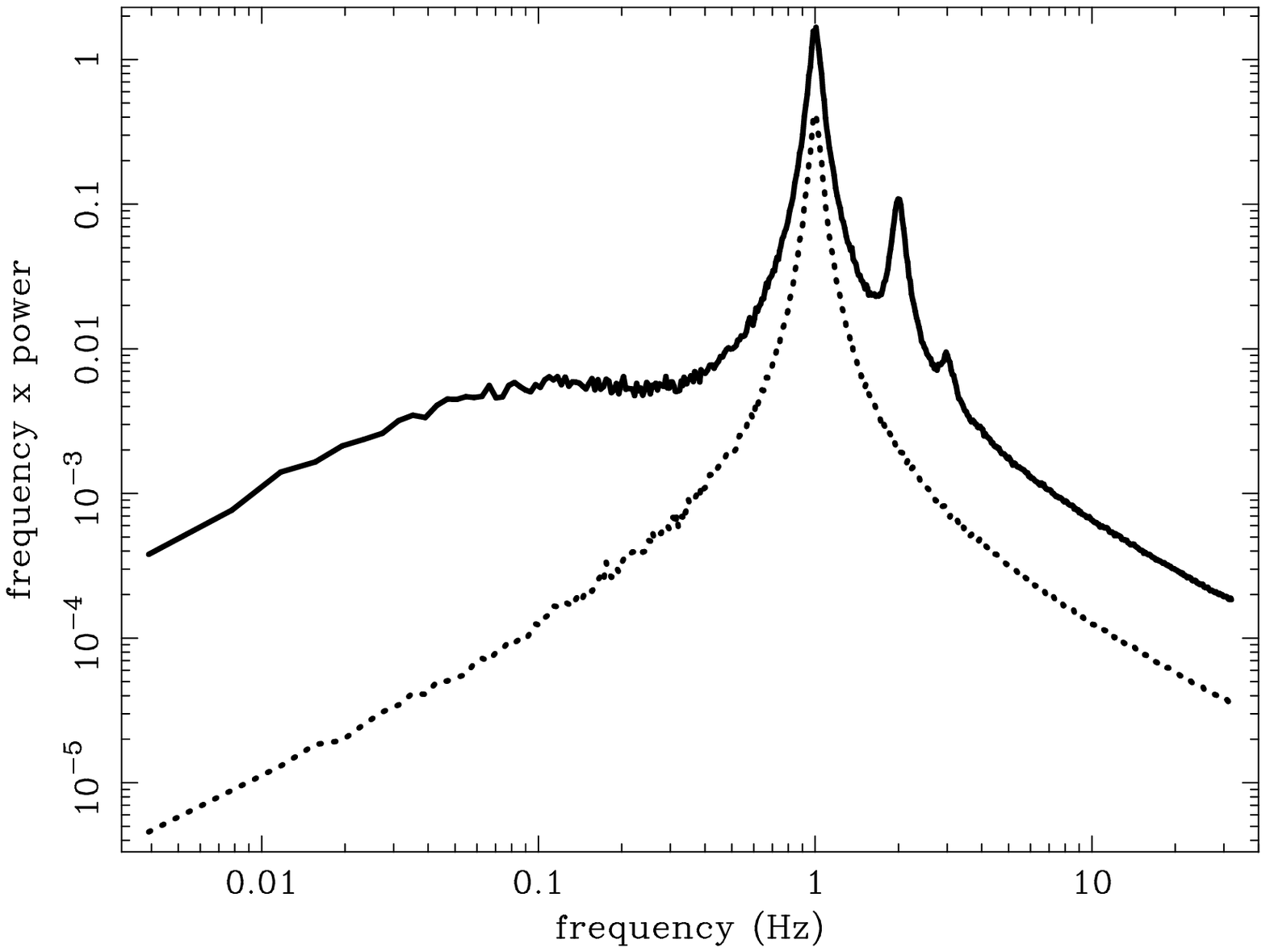}
}\caption{Solid line: PSD (plotted as frequency$\times$power)
of a simulated light curve made from an input light
curve with a sharp Lorentzian PSD (shown as the dotted line).} \label{explor}
\end{center}
\end{figure}

The situation is different, and quite interesting, for input light curves with PSDs containing
sharp features.  In Fig.~\ref{explor} we show the PSD produced by
applying the exponential transformation to a light curve produced by a
single sharp Lorentzian noise feature (i.e. a single QPO), with peak
frequency 1~Hz, quality factor (ratio between Lorentzian frequency and full width at 
half-maximum) $q=10$ and a fractional rms of 50~per~cent.  The multiple peaks are an
effect of the convolution of the PSD implicit in the exponential
transformation, where the large signal at the peak of the Lorentzian couples to itself and the
signals close to the peak, to produce the higher harmonics at separations of 1~Hz (in
theory there should be an infinite number of such harmonics, but the
3rd harmonic is only just visible, while even higher harmonics are lost in
the noise).  Interestingly, a shoulder is produced at around 0.1~Hz,
and in frequency$\times$power units, the flat top above the shoulder
corresponds to a $1/f$ power spectral shape, with slope 0 at lower
frequencies.  It is interesting that the exponential transformation of a
Lorentzian can produce what appears to be a rather broad continuum feature,
since the correlation between sharp QPO features and break frequencies
in continuum PSDs is already known in X-ray binary data \citep{wij99}. 
However, we must be careful not to read too much
into this interesting mathematical effect, because the simulated
Lorentzian peak has a much higher normalisation than typically observed,
and is still much larger than the `continuum' level.

\section[]{The calculation and interpretation of bicoherence}
\label{bicohapp}
The bicoherence $b^{2}$ is computed as follows (also see \citealt{mac02}
and references therein).  First the
light curve is split into $K$ segments, and the Fourier transform
$X$ calculated for each segment.  Then the bicoherence for the
pair of Fourier frequencies $k$, $l$ is calculated thus:
\begin{equation}
b^{2}(k,l)=\frac{|\sum_{i=1}^{K} X_{i}(k)X_{i}(l)X_{i}^{\ast}(k+l)|^{2}}
{\sum_{i=1}^{K} |X_{i}(k)X_{i}(l)|^{2}\:\sum_{i=1}^{K} |X_{i}(k+l)|^{2}}
\label{bicoheqn}
\end{equation}
The domain in which bicoherence measurements are independent of one
another is for bifrequencies $(k,\:l)$ where $l\leq k$ and $l+k\leq N/2$
where $N$ is the number of data points in the light curve segment (i.e. $N/2$
corresponds to the Nyquist frequency).
In data where the source variability is also contaminated with
additive Gaussian noise, the denominator to Equation~\ref{bicoheqn},
which we call $A_{\rm 1}$, must be
corrected for the effects of the noise (the noise cancels in the
numerator to leave only the source contribution).  The noise-corrected
denominator $A_{\rm 2}$ can be written as:
\begin{eqnarray}
A_{\rm 2} & = & \sum_{i=1}^{K} |X_{i}(k)X_{i}(l)|^{2}|-
n^{2}(|X_{i}(k)|^{2}+|X_{i}(l)|^{2}-n^{2}) 
\nonumber\\
& \times & \sum_{i=1}^{K} \left(|X_{i}(k+l)|^{2}-n^{2}\right)
\end{eqnarray}
where $n^{2}$ is the expected noise level of the PSD due to the
additive Gaussian process.  The derivation of this equation is similar to that of
the noise correction to the related
{\it coherence function} \citep{vau97,now99}, which is used to examine correlations between light curve 
phases at {\it identical} temporal frequencies,
measured in two different energy bands (unlike bicoherence which measures correlations
between phases at different frequencies in the same light curve).

Since the resulting bicoherence measure is forced to lie between 0 and 1, a bias
of $1/K$ must be subtracted from the bicoherence {\it before}
correcting for noise, i.e. calculate bicoherence using
Equation~\ref{bicoheqn}, then subtract $1/K$ from all measurements,
and finally multiply each measurement by the ratio of denominators
$A_{\rm 1}/A_{\rm 2}$.  {\it While this noise correction
is effective in accounting for additive Gaussian noise, it should be noted
that it is not strictly applicable for noise generated by Poisson
counting statistics}.  The reason for this is that the amplitude of
Poisson noise is correlated with $\sqrt{flux}$, so that the counting noise
variations are in fact coupled to the flux variations (albeit more
weakly than the intrinsic coupling between flux variations associated
with the linear rms-flux relation).  Therefore not only does the noise
in real photon counting data affect the denominator of the bicoherence
equation, it also adds to the numerator, to create a source of
spurious bicoherence.

Note that the numerator of the bicoherence equation, proportional to the modulus
squared of the `bispectrum', averages to zero in the case of aperiodic
variability with no coupling between variations on different
time-scales (i.e. at different frequencies), because the phases
of the signal at different Fourier frequencies are uncorrelated.  In
contrast, in the simple case where a pair of sine waves at frequencies $\nu_{1}$ and
$\nu_{2}$ are multiplied together, the coupling of the sine waves will produce a higher frequency
signal at $\nu=\nu_{1}+\nu_{2}$, which has a phase equal to the sum of phases of
the two sine waves and a signal power equal to the square of
the powers of the two sine waves and hence $b^{2}(\nu_{1},\nu_{2})=1$.
Thus, in simple terms, the bicoherence for a given pair of frequencies
$\nu_{1}$, $\nu_{2}$ indicates the fraction of the power at the
frequency $\nu_{1}+\nu_{2}$ which is produced by coupling of the
lower-frequency signals.  It is easy to see from the series expansion of the
exponential model (Appendix~\ref{nonlinproof}) why the model will produce light 
curves with a significant bicoherence.

\section[]{The rms-flux relations of other distributions}
\label{rmsdistrib}
\begin{figure}
 \par\centerline{\psfig{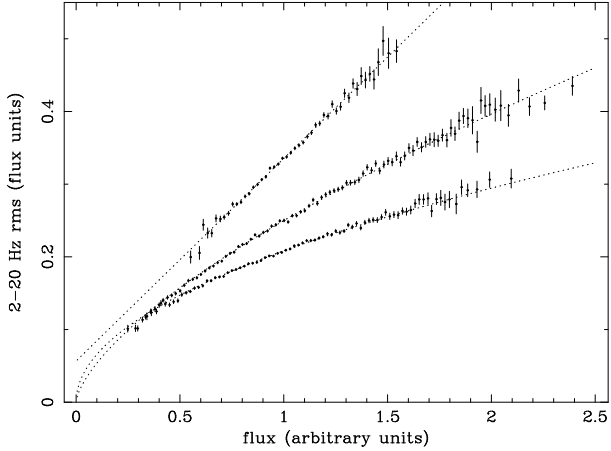}}
 \caption{\label{rmscompare} Comparison of rms-flux relations for different PSD shapes and
distributions $f[g(t)]$, where $g(t)$ is a Gaussian distribution of data.
From top to bottom: rms-flux relation for
$f[g(t)]=g(t)^{3}$ for a PSD slope of zero breaking to -2 above 2~Hz, 
fitted with a linear plus constant model (dotted line); $f[g(t)]=g(t)^{3}$ for a 
PSD similar to Cyg~X-1 in 
1996 Dec, fitted with a power-law of index 2/3 (dotted line); $f[g(t)]=g(t)^{2}$ for a 
PSD similar to Cyg~X-1 in 1996 Dec, fitted with a power-law of index 1/2 (dotted line).
See text for further details.
   }
\end{figure}

In order to see how general the linear rms-flux relation is, we can consider the rms-flux relations
of flux distributions other than lognormal.  It is a well known result in statistics that the values of
the sample variance and sample mean (i.e. estimates of mean and variance measured from a subset of 
an underlying population, such as a segment of a light curve) are independent if 
and only if the underlying population is Gaussian (e.g. \citealt{ken94}).  However, 
if the distribution
is not normal, e.g. it is skewed, then the sample mean and variance are not independent 
(with the degree
of correlation scaling with the skewness of the distribution, \citealt{kan85}), and hence the rms
from light curves with skewed distributions is correlated with flux (with the correlation having
the same sign as the skewness).  To determine the form of rms-flux relation resulting
from different distributions, we
can consider the simple case where the variance is small compared to changes in mean flux. 
Consider a non-Gaussian light curve $x(t)$ produced by some transformation of an underlying
Gaussian light curve, $g(t)$, i.e. $x(t)=f[g(t)]$.  For small changes in the flux, we can relate the 
variance of the Gaussian light curve $\delta g^{2}$ (which 
is independent of flux, and can be treated as constant in the binned rms-flux relation),
to the variance of the non-Gaussian light curve
$\delta x^{2}$ using the error equation (see also \citealt{bar47}):
\begin{equation}
\delta x^{2}= \left(\frac{\partial x(t)}{\partial g(t)}\right)^{2} \delta g^{2} .
\end{equation}
Taking the square-root of both sides then gives an approximate expression for the rms of $x(t)$,
$\sigma_{x}$.  For example, in the case of a lognormal distribution, $x(t)=\exp[g(t)]$ and hence
$\sigma_{x}\propto x(t)$, as expected.  However, if $x(t)$ is produced by the power-law transformation
of $g(t)$, $x(t)=g(t)^{\alpha}$, the relation becomes $\sigma_{x} \propto x(t)^{(1-\alpha^{-1})}$, i.e.
the rms-flux relation has a power-law form.  It should be stressed that this result only applies in
the case where $\sigma_{x}$ is small compared to $x(t)$, i.e. the fractional rms is small. This is
generally true in the case where the rms is measured at frequencies
significantly above any low-frequency break in the PSD
(below which PSD slope is zero), but may 
not be true where rms is measured close to the low-frequency 
break, in the regime where the data is white-noise, where the rms-flux
relation can become close to linear for a wide range of skewed flux distributions.  We demonstrate
this effect in Fig.~\ref{rmscompare}, which shows the rms-flux relations measured in the 2-20~Hz
range, for simulated light curves with a distribution $x(t)=g(t)^{\alpha}$, where $\alpha=3$ and
the PSD has either the same shape as seen in Cyg~X-1 in Dec 1996 (Fig.~\ref{dec1696psd}), or
is a singly broken power-law with a slope -2 above 2~Hz and slope zero below the break (hence the 
measured rms is a large fraction of the total rms and the mean flux).
The figure shows that the data where measured rms is small follows the predicted rms-flux relation
$\sigma_{x}\propto x(t)^{2/3}$, whereas the data with large measured rms shows
an approximately linear rms-flux relation (including a small constant offset on the rms axis).
We also show data for the Cyg~X-1-like PSD corresponding to a flux distribution with index $\alpha=2$,
which is well-fitted by the predicted rms-flux relation $\sigma_{x}\propto x(t)^{1/2}$.
These results demonstrate that, for the kind of rms-flux relations we measure using rms at relatively
high frequencies (e.g. 2-20~Hz
in Cyg~X-1 and equivalently high frequencies in AGN), 
the shape of the rms-flux relation is a good predictor
of the functional form of the underlying skewed distribution.  However, care must be taken when
measuring large values of rms, e.g. 
close to the low-frequency break, since in that case, approximately linear rms-flux relations
can be made for a variety of underlying distributions.

\bsp

\begin{thebibliography}{99}
\bibitem[\protect\citeauthoryear{Aitchison \& Brown}
{1957}]{ait57} Aitchison J., Brown J. A. C., 1957, The Lognormal Distribution,
Cambridge University Press, Cambridge
\bibitem[\protect\citeauthoryear{Armitage \& Reynolds}
{2003}]{arm03} Armitage P. J., Reynolds C. S., 2003, MNRAS, 341, 1041
\bibitem[\protect\citeauthoryear{Bak, Tang \& Wiesenfeld}
{1988}]{bak88} Bak P., Tang C., Wiesenfeld K., 1988, Phys. Rev. A, 38, 364
\bibitem[\protect\citeauthoryear{Bartlett}
{1947}]{bar47} Bartlett M. S., 1947, Biometrics, 3, 39
\bibitem[\protect\citeauthoryear{Belloni \& Hasinger}
{1990}]{bel90} Belloni T., Hasinger G., 1990, A\&A, 227, L33
\bibitem[\protect\citeauthoryear{Belloni, Psaltis \& van der Klis}
{2002}]{bel02} Belloni T., Psaltis D., van der Klis M., 2002, ApJ, 572, 392
\bibitem[\protect\citeauthoryear{Boller et al.}
{1997}]{bol97} Boller~Th., Brandt~W.~N., Fabian~A.~C., Fink~H.~H., 1997,
MNRAS, 289, 393
\bibitem[\protect\citeauthoryear{Box \& Cox}
{1964}]{box64} Box G. E. P., Cox D. R., 1964, J. R. Statistical Soc. B, 26, 211
\bibitem[\protect\citeauthoryear{Brandt et al.}
{1999}]{bra99} Brandt~W.~N., Boller~Th., Fabian~A.~C., Ruszkowski~M., 1999,
MNRAS, 303, L53
7
\bibitem[\protect\citeauthoryear{Christensen, Fogedby \& Jensen}
{1991}]{chr91} Christensen K., Fogedby H. C., Jensen H. J., 1991,
J. Sta. Phys., 63, 653
\bibitem[\protect\citeauthoryear{Christensen, Olami \& Bak}
{1992}]{chr92} Christensen K., Olami Z., Bak P., 1992,
Phys. Rev. Lett., 68, 241
\bibitem[\protect\citeauthoryear{Churazov, Gilfanov \& Revnivtsev}
{2001}]{chu01} Churazov~E., Gilfanov~M., Revnivtsev~M., 2001, MNRAS,
321, 759
\bibitem[\protect\citeauthoryear{Crow \& Shimizu}
{1988}]{cro88} Crow E. L., Shimizu K., 1988,
Lognormal Distributions: Theory and Applications, Dekker, New York
\bibitem[\protect\citeauthoryear{Dewangan et al.}
{2002}]{dew02} Dewangan G. C., Boller Th., Singh K. P., Leighly~K.~M.,
2002, A\&A, 390, 65
\bibitem[\protect\citeauthoryear{Doi}
{1978}]{doi78} Doi K., 1978, Nature, 275, 197
\bibitem[\protect\citeauthoryear{Done \& Gierlinski}
{2003}]{don03} Done C., Gierlinski M., 2003, MNRAS, 342, 1041
\bibitem[\protect\citeauthoryear{Edelson et al.}
{2002}]{ede02} Edelson~R., Turner~T.~J., Pounds~K., Vaughan~S.,
Markowitz~A., Marshall~H., Dobbie~P., Warwick~R., 2002, ApJ, 568, 610
\bibitem[\protect\citeauthoryear{Fabian}
{1979}]{fab79} Fabian A. C., 1979, Proc. R. Soc. London, Ser. A, 366, 449
\bibitem[\protect\citeauthoryear{Fackrell}
{1996}]{fac96} Fackrell J., 1996, PhD Thesis, Univ. Edinburgh
\bibitem[\protect\citeauthoryear{Gaskell}
{2003}]{gas03} Gaskell C. M., 2004, ApJ, 612, L21
\bibitem[\protect\citeauthoryear{Gierlinski \& Zdziarski}
{2003}]{gie03} Gierlinski M., Zdziarski~A.~A., 2003, MNRAS, 343, L84 (GZ03)
\bibitem[\protect\citeauthoryear{Gilfanov, Revnivtsev \& Molkov}
{2003}]{gil03} Gilfanov~M., Revnivtsev~M., Molkov~S., 2003, A\&A, 410, 217
\bibitem[\protect\citeauthoryear{Gleissner et al.}
{2004}]{gle04} Gleissner T., Wilms J., Pottschmidt K., Uttley P.,
Nowak~M.~A., Staubert~R., 2004, A\&A, 414, 1091
\bibitem[\protect\citeauthoryear{Gliozzi et al.}
{2002}]{gli02} Gliozzi M., Brinkmann W., R\"{a}th C., Papadakis I. E.,
Negoro~H., Scheingraber~H., 2002, A\&A 391, 875
\bibitem[\protect\citeauthoryear{Greb \& Rusbridge }
{1988}]{gre88} Greb U., Rusbridge M. G., 1988, Plasma Phys. \& Controlled Fusion, 30, 537
\bibitem[\protect\citeauthoryear{Green, M$^{\rm c}$Hardy \& Done }
{1999}]{gre99} Green A. R., M$^{\rm c}$Hardy I. M., Done C., 1999,
MNRAS, 305, 309
\bibitem[\protect\citeauthoryear{Hinich}
{1982}]{hin82} Hinich M. J., 1982, J. of Time Series Analysis, 3, 169
\bibitem[\protect\citeauthoryear{Kang \& Goldsman}
{1985}]{kan85} Kang K., Goldsman, D., 1985, in Gantz, D., Blais, G., Solomon, S., eds, Proc. of the 17th 
Conference on Winter Simulation, ACM Press, New York, USA, p. 211
ESA, Noordwijk, the Netherlands, p. 499
\bibitem[\protect\citeauthoryear{Kantz \& Schreiber}
{1997}]{kan97} Kantz H., Schreiber T., 1997, Nonlinear Time Series Analysis, Cambridge
University Press, Cambridge
\bibitem[\protect\citeauthoryear{Kendall}
{1994}]{ken94} Kendall, M. G., 1994, Kendall's Advanced Theory of Statistics 6th ed. v. 1, Edward Arnold, London
\bibitem[\protect\citeauthoryear{King et al.}
{2004}]{kin04} King A. R., Pringle J. E., West R. G., Livio~M., 2004,
MNRAS, 348, 111
\bibitem[\protect\citeauthoryear{Kotov, Churazov \& Gilfanov}
{2001}]{kot01} Kotov~O., Churazov~E., Gilfanov~M., 2001, MNRAS, 327, 799
\bibitem[\protect\citeauthoryear{Lehto}
{1989}]{leh89} Lehto H., 1989, in Hunt J., Battrick B., eds, Two Topics in 
X-ray Astronomy, ESA SP296. ESA, Noordwijk, the Netherlands, p. 499
\bibitem[\protect\citeauthoryear{Leighly}
{1999}]{lei99} Leighly~K. M., 1999, ApJS, 125, 297
\bibitem[\protect\citeauthoryear{Leighly \& O'Brien}
{1997}]{lei97} Leighly~K. M., O'Brien P. T., 1997, ApJ, 481, L15
\bibitem[\protect\citeauthoryear{Lochner, Swank \& Szymkowiak}
{1989}]{loc89} Lochner~J.~C., Swank~J.~H., Szymkowiak~A.~E., 1989, ApJ, 337, 823
\bibitem[\protect\citeauthoryear{Lochner, Swank \& Szymkowiak}
{1989}]{loc91} Lochner~J.~C., Swank~J.~H., Szymkowiak~A.~E., 1991, ApJ,
376, 295
\bibitem[\protect\citeauthoryear{Lyubarskii}
{1997}]{lyu97} Lyubarskii~Yu.~E., 1997, MNRAS, 292, 679
\bibitem[\protect\citeauthoryear{Maccarone \& Coppi}
{2002}]{mac02} Maccarone T. J., Coppi P. S., 2002, MNRAS, 336, 817 (MC02)
\bibitem[\protect\citeauthoryear{McClintock \& Remillard}
{2003}]{mcc03} McClintock J. E., Remillard R. A., 2003, to appear in W. H. G.
Lewin, M. van der Klis eds., Compact Stellar X-ray Sources, 
Cambridge University Press, Cambridge (astro-ph/0306213)
\bibitem[\protect\citeauthoryear{M$^{\rm c}$Hardy}
{1988}]{mch88} M$^{\rm c}$Hardy I. M., 1988, Mem. Soc. Astron. Ital.,
59, 239
\bibitem[\protect\citeauthoryear{M$^{\rm c}$Hardy et al.}
{2004}]{mch04} M$^{\rm c}$Hardy I. M., Papadakis~I.~E., Uttley~P.,
Page~M.~J., Mason~K.~O., 2004, MNRAS, 348, 783
\bibitem[\protect\citeauthoryear{Machida \& Matsumoto}
{2003}]{mac03} Machida M., Matsumoto R., 2003, ApJ, 585, 429
\bibitem[\protect\citeauthoryear{Malzac, Merloni \& Fabian}
{2004}]{mal04} Malzac J., Merloni A., Fabian A. C., 2004, MNRAS, 351, 253
\bibitem[\protect\citeauthoryear{Markowitz et al.}
{2003}]{mar03} Markowitz A. et al., 2003, ApJ, 593, 96
\bibitem[\protect\citeauthoryear{Merloni \& Fabian}
{2001}]{mer01} Merloni A., Fabian A. C., 2001, MNRAS, 328, 958
\bibitem[\protect\citeauthoryear{Mineshige et al.}
{1994}]{min94} Mineshige S., Ouchi N. B., Nishimori H., 1994, PASJ, 46, 97
\bibitem[\protect\citeauthoryear{Misra et al.}
{2004}]{mis04} Misra R., Harikrishnan K. P., Mukhopadhyay B., Ambika~G., 
Kembhavi~A.~K., 2004, ApJ, 609, 313
\bibitem[\protect\citeauthoryear{Nandra \& Papadakis}
{2001}]{nan01} Nandra K., Papadakis I. E., 2001, ApJ, 554, 710
\bibitem[\protect\citeauthoryear{Negoro \& Mineshige}
{2002}]{neg02} Negoro H., Mineshige S., 2002, PASJ, 54, L69
\bibitem[\protect\citeauthoryear{Nowak}
{2000}]{now00} Nowak M. A., 2000, MNRAS, 318, 361
\bibitem[\protect\citeauthoryear{Nowak et al.}
{1999}]{now99} Nowak M. A., Vaughan B. A., Wilms~J.,
Dove~J.~B., Begelman~M.~C., 1999, ApJ, 510, 874
\bibitem[\protect\citeauthoryear{Pottschmidt et al.}
{2003}]{pot03} Pottschmidt K. et al., 2003, A\&A, 407, 1039
\bibitem[\protect\citeauthoryear{Poutanen \& Fabian}
{1999}]{pou99} Poutanen~J., Fabian~A.~C., 1999, MNRAS, 306, L31
\bibitem[\protect\citeauthoryear{Poutanen \& Gierlinski}
{2003}]{pou03} Poutanen~J., Gierlinski M., 2003, MNRAS, 343, 1301
\bibitem[\protect\citeauthoryear{Priestley}
{1982}]{pri82} Priestley~M.~B., 1982, Spectral Analysis and Time
Series, Academic Press, London
\bibitem[\protect\citeauthoryear{Quilligan et al.}
{2002}]{qui02} Quilligan~F., McBreen B., Hanlon~L., McBreen S.,
Hurley~K.~J., Watson~D., 2002, A\&A, 385, 377
\bibitem[\protect\citeauthoryear{Rial \& Anaclerio}
{2000}]{ria00} Rial J. A., Anaclerio C. A., 2000, Quaternary Sci. Rev., 19, 1709
\bibitem[\protect\citeauthoryear{Scargle}
{1997}]{sca97} Scargle~J.~D., 1997, in G.~J. Babu, E.~D. Feigelson eds.,
Statistical Challenges in Modern Astronomy II, Springer-Verlag
(New York), p317
\bibitem[\protect\citeauthoryear{Subba Rao \& Gabr}
{1980}]{sub80} Subba Rao T., Gabr M. M., 1980, J. of Time Series Analysis, 1, 145
\bibitem[\protect\citeauthoryear{Takeuchi, Mineshige \& Negoro}
{1995}]{tak95} Takeuchi M., Mineshige S., Negoro H., 1995, PASJ, 47, 617
\bibitem[\protect\citeauthoryear{Taylor, Uttley \& M$^{\rm c}$Hardy}
{2003}]{tay03} Taylor R. D., Uttley P., M$^{\rm c}$Hardy I. M., 2003, MNRAS, 342, L31
\bibitem[\protect\citeauthoryear{Terrell}
{1972}]{ter72} Terrell~N.~J., 1972, ApJ, 174, L35
\bibitem[\protect\citeauthoryear{Theiler et al.}
{1992}]{the92} Theiler J., Eubank S., Longtin A., Galdrikian~B.,
Farmer~J.~D., 1992, Physica D, 58, 77
\bibitem[\protect\citeauthoryear{Theiler, Linsay \& Rubin}
{1994}]{the94} Theiler J., Linsay P.~S., Rubin~D.~M., 1994, in
Weigen~A.~S., Gerschenfeld~N.~A., eds, Santa Fe Institute Studies in the 
Sciences of Complexity, Proc. vol. 15, Addison-Wesley, p.~429
(http://arxiv.org/abs/comp-gas/9302003)
\bibitem[\protect\citeauthoryear{Timmer \& K\"{o}nig}
{1995}]{tim95} Timmer~J., K\"{o}nig~M., 1995, A\&A, 300, 707
\bibitem[\protect\citeauthoryear{Uttley}
{2004}]{utt04} Uttley P., 2004, MNRAS, 347, L61
\bibitem[\protect\citeauthoryear{Uttley \& M$^{\rm c}$Hardy}
{2001}]{utt01} Uttley~P., M$^{\rm c}$Hardy~I.~M., 2001, MNRAS, 323, L26
\bibitem[\protect\citeauthoryear{Uttley et al.}
{1999}]{utt99} Uttley~P., M$^{\rm c}$Hardy~I.~M., Papadakis~I.~E.,
Guainazzi~M., Fruscione~A., 1999, MNRAS, 307, L6
\bibitem[\protect\citeauthoryear{Uttley, M$^{\rm c}$Hardy \& Papadakis}
{2002}]{utt02} Uttley~P., M$^{\rm c}$Hardy~I.~M., Papadakis~I.~E.,
2002, MNRAS, 332, 231
\bibitem[\protect\citeauthoryear{Uttley et al.}
{2003}]{utt03} Uttley P., Fruscione A., M$^{\rm c}$Hardy I. M., Lamer~G., 2003, ApJ, 595, 656
\bibitem[\protect\citeauthoryear{Uttley et al.}
{2004}]{utt04b} Uttley P., Taylor R. D., M$^{\rm c}$Hardy I. M., Page~M.~J., Mason~K.~O.,
Lamer~G., Fruscione~A., 2004, MNRAS, 347, 1345
\bibitem[\protect\citeauthoryear{van der Klis}
{1995}]{vdk95} van der Klis M., 1995, in Lewin~H.~G., Van Paradijs~J., van
den Heuvel~P.~J., eds, X-Ray Binaries. Cambridge Univ. Press (Cambridge),
p. 252
\bibitem[\protect\citeauthoryear{Vaughan et al.}
{2003}]{vau03b} Vaughan S., Edelson R., Warwick~R.~S.,
Uttley~P., 2003, MNRAS, 345, 1271
\bibitem[\protect\citeauthoryear{Vaughan, Fabian \& Nandra}
{2003}]{vau03a} Vaughan S., Fabian A. C., Nandra K., 2003, MNRAS, 339, 1237
\bibitem[\protect\citeauthoryear{Vaughan \& Nowak}
{1997}]{vau97} Vaughan B. A., Nowak M. A., 1997, ApJ, 474, L43
\bibitem[\protect\citeauthoryear{Wijnands \& van der Klis}
{1999}]{wij99} Wijnands R., van der Klis M., 1999, ApJ, 514, 939
\bibitem[\protect\citeauthoryear{\.{Z}ycki}
{2003}]{zyc03} \.{Z}ycki P. T., 2003, MNRAS, 340, 639
\end{thebibliography}
\end{document}